\documentclass[11pt,twoside]{article}

\newcommand{\boldsection}[1]{\section{\boldmath#1}}
\newcommand{\boldsubsection}[1]{\subsection[#1]{\boldmath#1}}
\newcommand{\boldsubsubsection}[1]{\subsubsection[#1]{\boldmath#1}}

\makeatletter 
\let\mv@BOLD=\mv@normal

\renewcommand{\pmod}[1]{\allowbreak\mskip12mu({\operator@font mod}\,\,#1)}

\newcommand{\Tr}{\mathop{\operator@font Tr}\nolimits}
\newcommand{\Ad}{\mathop{\operator@font Ad}\nolimits}
\newcommand{\Id}{\mathop{\operator@font Id}\nolimits}

\newcommand{\Const}{\mathop{\operator@font const.}\nolimits}
\newcommand{\Dim}{\mathop{\operator@font dim}\nolimits}

\renewcommand{\Im}{\mathop{\operator@font Im}\nolimits}
\renewcommand{\Re}{\mathop{\operator@font Re}\nolimits}
\makeatother 

\newcommand{\dual}{{}^*\!}

\newcommand{\cA}{{\cal A}}
\newcommand{\cD}{{\cal D}}
\newcommand{\cH}{{\cal H}}
\newcommand{\cJ}{{\cal J}}

\newcommand{\bG}{\bar G}
\newcommand{\bH}{\bar H}
\newcommand{\bJ}{\bar J}
\newcommand{\bM}{\bar M}
\newcommand{\bP}{\bar P}
\newcommand{\bPhi}{\bar\Phi}
\newcommand{\bQ}{\bar Q}
\newcommand{\bT}{\bar T}
\newcommand{\bV}{\bar V}
\newcommand{\bY}{\bar Y}
\newcommand{\ba}{\bar a}
\newcommand{\bb}{\bar b}
\newcommand{\bd}{\bar d}
\newcommand{\bdelta}{\bar\delta}
\newcommand{\be}{\bar e}
\newcommand{\bepsilon}{\bar\epsilon}
\newcommand{\bg}{\bar g}
\newcommand{\bgamma}{\bar\gamma}
\newcommand{\bh}{\bar h}
\newcommand{\bk}{\bar k}
\newcommand{\bkappa}{\bar\kappa}
\newcommand{\bmu}{\bar\mu}
\newcommand{\bpi}{\bar\pi}
\newcommand{\bphi}{\bar\phi}
\newcommand{\brho}{\bar\rho}
\newcommand{\bt}{\bar t}
\newcommand{\btau}{\bar\tau}
\newcommand{\bu}{\bar u}
\newcommand{\bv}{\bar v}
\newcommand{\bvarphi}{\bar\varphi}

\newcommand{\hC}{\hat C}
\newcommand{\hJ}{\hat J}
\newcommand{\hM}{\hat M}
\newcommand{\hN}{\hat N}
\newcommand{\hP}{\hat P}
\newcommand{\hR}{\hat R}
\newcommand{\hV}{\hat V}
\newcommand{\hW}{\hat W}
\newcommand{\hg}{\hat g}
\newcommand{\hlambda}{\hat\lambda}
\newcommand{\hpi}{\hat\pi}
\newcommand{\hrho}{\hat\rho}
\newcommand{\hs}{\hat s}

\newcommand{\tdB}{\tilde B}
\newcommand{\tdG}{\tilde G}
\newcommand{\tdJ}{\tilde J}
\newcommand{\tdM}{\tilde M}
\newcommand{\tdN}{\tilde N}
\newcommand{\tdP}{\tilde P}
\newcommand{\tdR}{\tilde R}
\newcommand{\tdV}{\tilde V}
\newcommand{\tdW}{\tilde W}
\newcommand{\tdg}{\tilde g}
\newcommand{\tdh}{\tilde h}
\newcommand{\tdlambda}{\tilde\lambda}
\newcommand{\tdmu}{\tilde\mu}
\newcommand{\tdnu}{\tilde\nu}
\newcommand{\tdpi}{\tilde\pi}
\newcommand{\tdrho}{\tilde\rho}
\newcommand{\tds}{\tilde s}
\newcommand{\tdtau}{\tilde\tau}
\newcommand{\tdy}{\tilde y}

\newcommand{\bcA}{\bar\cA}
\newcommand{\bcJ}{\bar\cJ}
\newcommand{\tdcA}{\tilde\cA}
\newcommand{\tdcJ}{\tilde\cJ}

\newcommand{\bfE}{{\bf1}}

\newcommand{\bfA}{{\bf A}}
\newcommand{\bfC}{{\bf C}}
\newcommand{\bfR}{{\bf R}}
\newcommand{\bfS}{{\bf S}}

\newcommand{\bfalpha}{\hbox{\boldmath$\alpha$}}
\newcommand{\bfbeta}{\hbox{\boldmath$\beta$}}

\begin{document}

\hbox to\hsize{%
	\hfill MPI-PhT 98-42}
\hbox to\hsize{%
	\hfill May 1998}
\vspace{5mm}

\begin{center}
{\LARGE\bf\boldmath On Nonlinear $\sigma$-Models\\
	arising in (Super-)Gravity

}
%
\vspace{5mm}

Peter Breitenlohner and Dieter Maison\\
Max-Planck-Institut f\"ur Physik\\
(Werner-Heisenberg-Institut)\\
F\"ohringer Ring 6, D-80805 M\"unchen, Germany
\end{center}
\vspace{5mm}

\section*{Abstract}
In a previous paper with Gibbons~\cite{BMG} we derived a list of three
dimensional symmetric space $\sigma$-model obtained by dimensional reduction
of a class of four dimensional gravity theories with abelian gauge fields
and scalars. Here we give a detailed analysis of their group theoretical
structure leading to an abstract parametrization in terms of `triangular'
group elements. This allows for a uniform treatment of all these models.
As an interesting application we give a simple derivation of a `Quadratic
Mass Formula' for strictly stationary black holes.


\section{Introduction}

Starting from the maximal eleven dimensional supergravity theory many
different models with $N\geq 1$ supersymmetries have been constructed in
lower dimensions via Ka{\l}uza-Klein reductions. In the case of
compactification on a torus the use of adapted coordinates leads to
abelian gauge symmetries derived from the general covariance in
eleven dimensions. Depending on the dimension some of the abelian vector
resp.\ tensor gauge fields can be replaced by scalar potentials via Hodge
duality. Experience shows that all the scalars of these models organize
themselves into the simple structure of a non-linear sigma model with a
coset space $G/H$ as target space. Even more, these coset spaces have the
structure of a Riemannian resp.\ pseudo-Riemannian non-compact symmetric
space. As the number of scalars increases when the dimension is reduced it
reaches a maximum in three dimensions. Although further reduction to two
dimensions results in another scalar from the 3-metric, the latter plays
a different role and does not lead to a further increase of
$G/H$.\footnote{There is however a dramatic increase of $G$ to the
infinite dimensional Geroch group, if the complete integrability of the two
dimensional theories is taken into account~\cite{Julia,BM}.}
In our previous paper with Gibbons~\cite{BMG} we gave a general
classification of four dimensional bosonic theories
leading to three dimensional reductions allowing for Ehlers-Harrison type
transformations. Besides the models obtained from supergravities we found
many new cases including certain infinite series, all of them resulting in
symmetric space sigma models.
The present paper is devoted to an elaboration of the specific structure
of all these coset spaces.

An essential step for the identification of the particular sigma model
corresponding to a theory is the parametrization of the coset space
with the scalar fields of the theory. Considering specific examples one
finds a typical structure of the target space metric. While some of the
fields appear only in polynomial form others do not. This has been exploited
in the past to construct certain triangular matrix representations of the
coset spaces, such that the polynomial dependence is related to the
nilpotent (off-diagonal) part of the matrices. This reminds very much of the
Iwasawa decomposition for semi-simple Lie groups resp.\ its reduction to the
coset space. There is a corresponding structure in the action of
the various transformations of $G$ on the fields, in particular the
generalized `Ehlers' and `Harrison' transformations. In fact, this will
be the starting point of our structural analysis of the Lie algebra $g$ of
$G$. It turns out that all the different cases can be characterized by
certain matrices made up from the strucure constants of $g$. As a result,
we not only get the complete action of $g$ on the fields, but also obtain an
abstract parametrization of the coset space $G/H$ in terms of the fields via
the exponential map. Clearly, in order to obtain a concrete parametrization,
the above mentioned matrices have to be worked out for each case resp.\
family of models individually and we shall exemplify this in detail.
Nevertheless for certain applications our knowledge on the general structure
of $G$ is sufficient without the need of any specific parametrization.
A particularly interesting case is provided by the `Quadratic Mass
Formula' for strictly stationary single black hole solutions, generalizing
work of Heusler~\cite{Heusler}. Another important application we have in mind
but not yet completely worked out is a proof of the uniqueness of the
`Generalized Kerr Solutions', which we announced already in~\cite{BMG}.

In Chapter~\ref{c_redu} we review the dimensional reduction from four to
three dimensions for stationary or axisymmetric solutions, concentrating
on those cases, where the scalars of the 3-dimensional theory form a
nonlinear $\sigma$-model, i.e.\ parametrize a pseudo Riemannian or Riemannian
symmetric space and consequently there exist generalized Ehlers-Harrison
transformations. We closely follow the treatment in~\cite{BMG} and for the
convenience of the reader we repeat the list of all such cases from that paper.
In Chapter~\ref{c_lie} we analyze the general structure of the Lie algebra
$g$ and derive consistency conditions for the coefficients used to parametrize
the structure constants of the Lie algebra $g$ for all the possible cases.
In Chapter~\ref{c_sigma} we discuss the parametrization of coset representatives
forming a `triangular' subgroup $T\subset G$, the action of (infinitesimal)
Ehlers-Harrison transformations, and the form of the Lagrangian built from the
Lie algebra valued conserved current $J$. Chapter~\ref{c_black} contains
two simple applications for strictly stationary black holes.
In the appendix we give a detailed discussion of all the cases.

\section{Dimensional Reduction from 4 to 3 Dimensions}
\label{c_redu}

We consider 4-dimensional theories with scalars forming a nonlinear
$\sigma$-model coupled to gravity and $k$ abelian gauge fields.
Stationary solutions of the field equations can be described in terms of
a `dimensionally reduced' 3-dimensional theory. We will closely follow the
analysis in~\cite{BMG} and use a rather similar notation.

\subsection{The 4-Dimensional Theory}

We start from a Lagrangian field theory over a space-time manifold
$\Sigma_4$ with coordinates $x^a$ and metric $g_{ab}(x)$ (with
$ds^2=g_{ab}dx^adx^b>0$ along time-like directions). Let $\bPhi$ be a
Riemannian symmetric space with (real) coordinates $\bphi^i$ and metric
$\bgamma_{ij}(\bphi)$. The nonlinear $\sigma$-model with target space
$\bPhi$ is characterized by the kinetic term ${1\over2}g^{ab}(x)
\partial_a\bphi^i(x)\partial_b\bphi^j(x)\bgamma_{ij}(\bphi(x))$.
We are interested in particular in the case $\bPhi=\bG/\bH$, where $\bG$ is a
noncompact Lie group with maximal compact subgroup $\bH$. We choose a basis
$s_i$, $i=1,\ldots,\Dim\bG$ for the Lie algebra $\bg$ of $\bG$ with
commutators
$$
[s_i,s_j]=f_{ij}{}^ls_l\;.
$$
Due to the assumption that $\bG/\bH$ is a symmetric space, there exists an
involutive automorphism $\btau$ of $\bg$
$$
\btau([X,Y])=[\btau(X),\btau(Y)]\quad
{\rm for\ all\ } X,Y\in\bg\;,\qquad
{\btau}^2=\Id\;,
$$
such that
$$
\bh=\{X\in\bg:\btau(X)=X\}\;.
$$

We can describe elements of the coset space $\bG/\bH$ by coset
representatives $\bpi(x)\in\bG$. The group $\bG$ acts on these
representatives, but in addition there is the gauge group $\bH$ changing
the coset representatives
$$
\bpi(x)\mapsto\bv(x)\bpi(x)\bu^{-1}\;,\quad{\rm with}\quad
\bu\in\bG\;,\quad\bv(x)\in H\;.
$$
In order to eliminate the gauge degrees of freedom we consider the gauge
invariant group element $\bmu=\btau(\bpi^{-1})\bpi$ and the Lie algebra
valued currents $\bJ=\bJ_adx^a={1\over2}\bmu^{-1}d\bmu$. This allows us to
rewrite the kinetic term for the scalars in the form
${1\over4}<\bJ_a,\bJ^a>_{\bg}$, where $<\cdot,\cdot>_{\bg}$ is a
suitable (not necessarily unique) invariant scalar product on the Lie
algebra $\bg$. Finally, let there be $k$ (real) abelian vector fields
(`electric' potentials) $B_a=(B_a^I)$, $I=1,\ldots,k$ with field strengths
$G_{ab}=\partial_aB_b-\partial_bB_a$ and their duals $\dual G_{ab}$.

For the coupled system of gravity, scalars, and vector fields we choose
the action
$$
S_4=\int\limits_{\Sigma_4}\sqrt{|g|}d^4x\left(
  -{1\over2}R+{1\over4}<\bJ_a,\bJ^a>_{\bg}
  -{c\over8}G^T_{ab}(\tdmu G^{ab}-\tdnu\,\dual G^{ab})\right)\;,
$$
where $R$ is the scalar curvature, $c$ is a positive constant,
$\tdmu(\bphi)$ and $\tdnu(\bphi)$ are symmetric $k\times k$ matrices, and
$\tdmu$ is positive definite. The constant $c$ could be absorbed by
rescaling the matrices $\tdmu$ and $\tdnu$ and\slash or the vector fields
$B_a$. Rescaling the matrices would, however, invalidate the identification
of $\tdmu$ and $\bM$ (defined below) as matrix representatives of elements
of the group $\bG$ (compare Chapter~\ref{c_sigma}). Rescaling the vector
fields may be equally undesirable; for the Ka{\l}uza-Klein theories
discussed in detail in Section~\ref{s_sl} the vector fields $B_a^I$ are
directly related to components of the metric in $4+n$ dimensions. For
the generalized Einstein-Maxwell theories discussed in Section~\ref{s_su}
one would like to normalize the vector fields, and thus the charges, such
that a Reissner-Nordstr{\o}m black hole with charge $q$ has mass $m\ge|q|$.

The field equations for the vector fields
$\nabla_a(\tdmu G^{ab}-\tdnu\,\dual G^{ab})=0$ can be interpreted as Bianchi
identitities for field strengths $H_{ab}=\partial_aC_b-\partial_bC_a$
derived from `magnetic' potentials $C_a$. Choosing
$\dual H_{ab}=\eta(\tdmu G_{ab}-\tdnu\,\dual G_{ab})$ with some constant
orthogonal matrix $\eta$, we obtain $2k$ vector fields $A_a$ and field
strengths $F_{ab}=\partial_aA_b-\partial_bA_a$ satisfying the linear relation
$$
F_{ab}=\bY\bM\,\dual F_{ab}\;,\quad
F_{ab}=\left(\begin{array}{c}G_{ab}\\H_{ab}\end{array}\right)\;,\quad
A_m=\left(\begin{array}{c}B_m\\C_m\end{array}\right)\;,
$$
with matrices $\bM=\bM^T$ and $\bY=-\bY^T$ such that $\bY\bM\bY=-\bM^{-1}$
$$
\bY=\left(\begin{array}{cc}0&\eta^T\\-\eta&0\end{array}\right)\;,\quad
\bM=\left(\begin{array}{cc}\tdmu+\tdnu\tdmu^{-1}\tdnu&\tdnu\tdmu^{-1}\eta^T\\
    \eta\tdmu^{-1}\tdnu&\eta\tdmu^{-1}\eta^T\end{array}\right)\;.
$$
Assume there is a $2k$-dimensional real matrix representation $\brho$ of
$\bG$ with
$$
\brho:\bpi\mapsto\brho(\bpi)=\bP\;,\quad
\brho(\btau(\bpi))=\bP^{T-1}\;,\quad
\brho(\bmu)=\bP^T\bP=\bM\;.
$$
The field equations for the vectors are then $\bG$ invariant, provided
the action of $\bG$ on the field strenths is
$\bG\ni\bu:F_{ab}\mapsto\brho(\bu)F_{ab}$. Thus the group $\bG$ acts
nonlinearly on the scalars $\bphi$, but acts linearly on the field strengths
$G_{ab}$ and their duals $\dual G_{ab}$ (with coefficient depending on the
scalars). The contribution of the vector fields to the gravitational and
scalar field equations can be similarly expressed in terms of $F_{ab}$
in explicitly $\bG$-covariant form. This `on-shell' symmetry can, however,
in general not be formulated as an invariance of the action.

\subsection{The 3-Dimensional Theory}

A solution of the 4-dimensional field equations with a 1-parameter symmetry
group is characterized by a Killing vector field $K$ such that the Lie
derivative with respect to $K$ of the metric, scalars, and field
strengths vanishes. The solution is stationary, strictly stationary, or
axisymmetric if $K$ is asymptotically time-like, everywhere time-like, or
asymptotically space-like respectively. In the following we will mostly
assume that the solution is strictly stationary, i.e.\ $\Delta=K_aK^a>0$,
but the results hold as well for axisymmetric solutions with $\Delta<0$.

Although the dimensional reduction from~4 to~3 dimensions could be
formulated in a coordinate independent form, the discussion is simplified by
choosing adapted coordinates such that the isometry is just a translation
(e.g.\ $x^a=(x^m,t)$ with $K={\partial\over\partial t}$). The metric,
scalars, and field strengths will then depend only on the three coordinates
$x^m$, $m=1,2,3$ parametrizing the orbit space $\Sigma_3$ of the action of
$K$; in a suitable gauge this will also be true for the vector fields.
We decompose the metric and vector fields into pieces perpendicular and
parallel to $K_a=(\Delta k_m,\Delta)$
$$
g_{ab}=\left(\begin{array}{cc}
   -{1\over\Delta}h_{mn}+\Delta k_mk_n&\Delta k_n\\
   \Delta k_m&\Delta\end{array}\right)\;,\quad
B_a=\left(\begin{array}{c}\tdB_m+Bk_m\\B\end{array}\right)\;.
$$

Inserting this decomposition into the action $S_4$ and omitting the
integration over $dt$ as well as surface terms, we obtain the action $S_3$
describing a Lagrangian field theory over the orbit space $\Sigma_3$, the
dimensionally reduced theory. The action $\Sigma_3$ depends on the 3-vectors
$k_m$ and $\tdB_m$ only through their field strengths
$k_{mn}=\partial_mk_n-\partial_nk_m$ and
$\tdB_{mn}=\partial_m\tdB_n-\partial_n\tdB_m$; the resulting field
equations can again be interpreted as consistency conditions for the
existence of dual potentials. For $\dual\tdB_{mn}$ these are just the
parallel components $C$ of $C_a$, thus the $k$ vectors $B_a$ of the
4-dimensional theory yield $2k$ scalars $A$. The `twist' $\dual k_{mn}$ can
be expressed in terms of the `twist vector' $\omega_m$ with the `twist
potential' $\psi$
$$
A=\left(\begin{array}{c}B\\C\end{array}\right)\;,\quad
\omega_m=\partial_m\psi+{c\over2}A^T\bY^{-1}\partial_mA\;.
$$
Performing these dualizations via Lagrange multipliers finally yields the
dimensionally reduced action in the form
$$
\begin{array}{c}
\displaystyle
S_3=\int\limits_{\Sigma_3}\sqrt{|h|}d^3x\Bigl(
  {1\over2}R-{1\over4}<\bJ_m,\bJ^m>_{\bg}
  +{c\over4\Delta}\partial_mA^T\bM\,\partial^m\!A
\qquad\quad\\\displaystyle\qquad\quad
  -{1\over4\Delta^2}(\partial_m\Delta\,\partial^m\!\Delta+\omega_m\omega^m)
  \Bigr)\;,
\end{array}
$$
where the 3-metric $h_{mn}$ and its inverse are used to compute the scalar
curvature $R$ as well as to raise indices. We have used the rescaled
3-metric $h_{mn}$, assuming $\Delta\ne0$, in order to obtain an action
without additional coefficient for the scalar curvature.

The action $S_3$ describes a nonlinear $\sigma$-model with target space
$\Phi$ coupled to 3-dimensional gravity (note, however, that 3-dimensional
gravity has no dynamical degrees of freedom). $\Phi$ is a homogeneous space
parametrized by $(\Dim\bG-\Dim\bH)+2k+2$ scalars
$\phi=(\bphi,A,\Delta,\psi)$. The theory is invariant under various
transformation (with constant parameters):\\
(1)~twist gauge transformations $\psi\to\psi+\chi$,\\
(2)~`electromagnetic' gauge transformations $A\to A+\alpha$,
$\psi\to\psi+{c\over2}A^T\bY^{-1}\alpha$,\\
(3)~scale transformations $A\to\sqrt\zeta A$, $\Delta\to\zeta\Delta$,
$\psi\to\zeta\psi$, and\\
(4)~$\bG$ transformations acting on $\bphi$ and $A$.\\
The field equations for the scalar fields are equivalent to the
conservation laws for corresponding Noether currents.

The 3-metric $h_{mn}$ is positive definite for $\Delta>0$ (strict
stationarity) whereas $\Phi$ has an indefinite metric whith $2k$ negative
eigenvalues. For the axisymmetric case we will use primed variables
$\Phi'=(\bphi,A',\Delta',\psi')$ with $\Delta'<0$; the metric on $\Phi'$
is positive definite but the 3-metric $h'_{mn}$ has signature $(+,-,-)$.
The restriction of $\Phi$ (or $\Phi'$) to the submanifold $A=0$
(or $A'=0$) is the symmetric space $\bG/\bH\otimes SL(2)/SO(2)$, invariant
under the well known (infinitesimal) Ehlers transformation $\delta\bphi=0$,
$\delta\Delta=2\Delta\psi$, $\delta\psi=\psi^2-\Delta^2$~\cite{Neugebauer}.

For some such theories the Ehlers transformation can be extended to an
invariance of the whole target space $\Phi$ or $\Phi'$; commuting this
generalized Ehlers transformation with (infinitesimal) electromagnetic gauge
transformations finally yields generalized Harrison
transformations~\cite{Harrison}. All these transformations form a noncompact
Lie group $G$ with maximal compact
subgroup $H'$ and the target space is either the Riemannian symmetric space
$\Phi'=G/H'$ or the pseudo Riemannian symmetric space $\Phi=G/H$, where $H$
is a noncompact real form of $H'$. The dimensions of $G$ and $H$ are
$$
\Dim G=\Dim\bG+\Dim SL(2)+4k\;,\quad
\Dim H=\Dim\bH+\Dim SO(2)+2k\;.
$$
The dimensionally reduced action can then be expressed in the form
$$
S_3=\int\limits_{\Sigma_3}\sqrt{|h|}d^3x\left(
  {1\over2}R+L\right)\;,
\quad{\rm with}\quad
L=-{1\over4}<J_m,J^m>\;,
$$
where $<\cdot,\cdot>$ is a suitably normalized invariant scalar product
on the Lie Algebra $g$ of $G$.

The $G/H$ $\sigma$-model is characterized by two commuting involutive
automorphisms $\tau$ and $\tau'$: All elements of $G$ that are invariant
under $\tau$ form the subgroup $H$, all elements invariant under $\tau'$
form the subgroup $H'$, and all elements invariant under $\tau\tau'$ form
the subgroup $\bG\otimes SL(2)$. The restriction of $\tau$ or $\tau'$
to $\bG$ is the original automorphism $\btau$; the restriction
to $SL(2)$ is the automorphism defining the maximal compact subgroup $SO(2)$.
We choose a basis $t_i$ for the Lie algebra $sl(2)$, where $t_+=e$, $t_0=d$,
and $t_-=k$ satisfy the commutation relations
$$
[d,e]=e\;,\quad [d,k]=-k\;,\quad [e,k]=2d\;,
$$
and
$$
\tau(e)=-k\;,\quad \tau(d)=-d\;,\quad \tau(k)=-e\;.
$$

We may assume that the representation $\brho$ of $\bG$ is faithful, i.e.,
that all scalars of the 4-dimensional $\bG/\bH$ $\sigma$-model couple to the
vector fields. The group $G$ will then be simple, and Table~\ref{t_list} (a
reproduction of Table~2 in~\cite{BMG}) lists all possible cases.
\begin{table}[p]
\caption[t_list]{\label{t_list}List of all symmetric spaces obtained by
dimensional reduction from four to three dimensions of theories with scalars
and vectors (reproduced from Table~2 in \cite{BMG}).}
\begin{center}
\small
\newcommand{\D}[1]{$\displaystyle\mathop{#1}\limits^{}$}
\begin{tabular}{r|c|c|c|c|}
\hline
\#&\D{G/H}&\D{\bG/\bH}&\D{\Dim\bG/\bH}&\D{k}\\
\hline
1&\D{SL(n+2)/SO(n,2)}&\D{GL(n)/SO(n)}&\D{n(n+1)\over2}&$n$\\
2&\D{SU(p+1,q+1)\over S(U(p,1)\times U(1,q)}&
  \D{U(p,q)/(U(p)\times U(q))}&$2pq$&$p+q$\\
\noalign{\vskip-2.5ex}&&&&\\
3&\D{SO(p+2,q+2)\over SO(p,2)\times SO(2,q)}&
  \D{{SO(p,q)\over SO(p)\times SO(q)}\times{SO(2,1)\over SO(2)}}&
  $pq+2$&$p+q$\\
\noalign{\vskip-2.5ex}&&&&\\
4&\D{SO^*(2n+4)/U(n,2)}&
  \D{{SO^*(2n)\over U(n)}\times{SU(2)\over SU(2)}}&$n(n-1)$&$2n$\\
\noalign{\vskip-2.5ex}&&&&\\
5&\D{Sp(2n+2;\bfR)/U(n,1)}&\D{Sp(2n;\bfR)/U(n)}&$n(n+1)$&$n$\\
6&\D{G_{2(+2)}\over SU(1,1)\times SU(1,1)}&\D{SU(1,1)/U(1)}&2&2\\
\noalign{\vskip-2.5ex}&&&&\\
7&\D{F_{4(+4)}\over Sp(6;\bfR)\times SU(1,1)}&\D{Sp(6;\bfR)/U(3)}&12&7\\
\noalign{\vskip-2.5ex}&&&&\\
8&\D{E_{6(+6)}/Sp(8;\bfR)}&\D{SL(6)/SO(6)}&20&10\\
\noalign{\vskip-2.5ex}&&&&\\
9&\D{E_{6(-2)}\over SU(3,3)\times SU(1,1)}&
  \D{SU(3,3)\over S(U(3)\times U(3))}&18&10\\
\noalign{\vskip-2.5ex}&&&&\\
10&\D{E_{6(-14)}\over SO^*(10)\times SO(2)}&\D{SU(5,1)/U(5)}&10&10\\
11&\D{E_{7(+7)}/SU(4,4)}&\D{SO(6,6)\over SO(6)\times SO(6)}&36&16\\
12&\D{E_{7(-5)}\over SO^*(12)\times SO(2,1)}&\D{SO^*(12)/U(6)}&30&16\\
\noalign{\vskip-2.5ex}&&&&\\
13&\D{E_{7(-25)}\over E_{6(-14)}\times SO(2)}&
   \D{SO(10,2)\over SO(10)\times SO(2)}&20&16\\
\noalign{\vskip-2.5ex}&&&&\\
14&\D{E_{7(+8)}/SO^*(16)}&\D{E_{7(+7)}/SU(8)}&70&28\\
\noalign{\vskip-2.5ex}&&&&\\
15&\D{E_{8(-24)}\over E_{7(-25)}\times SU(1,1)}&
   \D{E_{7(-25)}\over E_{6(-78)}\times SO(2)}&54&28\\
\noalign{\vskip-2.0ex}&&&&\\
\hline
\end{tabular}
\end{center}
\end{table}

\section{Parametrization of the Lie Algebra}
\label{c_lie}

\subsection{Commutation Relations}

We choose a basis for the Lie algebra $g$ of $G$ consisting of the
generators $s_i$ of $\bg$, the generators $t_i$ of $sl(2)$, and $4k$
additional generators $h_i$, $a_i$, $i=1,\ldots,2k$ transforming as
doublet under $sl(2)$
$$
\begin{array}{lll}
\relax[d,h_i]={1\over2}h_i\;,&
\relax[e,h_i]=0\;,&
\relax[k,h_i]=-a_i\;,\\
\relax[d,a_i]=-{1\over2}a_i\;,&
\relax[e,a_i]=-h_i\;,&
\relax[k,a_i]=0\;,
\end{array}
$$
and with a $2k$-dimensional real matrix representation $\brho$ of $\bg$
$$
\brho:s_i\mapsto\brho(s_i)=R_i\;,\quad
[s_i,h\cdot\alpha]=h\cdot R_i\alpha\;,\quad
[s_i,a\cdot\alpha]=a\cdot R_i\alpha\;.
$$
The remaining commutators can be determined from the Jacobi identities
$J(x,y,z)\equiv[[X,Y],Z]+[[Y,Z],X]+[[Z,X],Y]=0$:
$$
\begin{array}{l}
J(h\cdot\alpha,h\cdot\beta,d)\Rightarrow
[h\cdot\alpha,h\cdot\beta]=\alpha\cdot y\beta\,e\;,
\\
J(h\cdot\alpha,h\cdot\beta,k)\Rightarrow
[h\cdot\alpha,a\cdot\beta]=\alpha\cdot y\beta\,d
  +\alpha\cdot x^l\beta\,s_l\;,
\\
J(h\cdot\alpha,a\cdot\beta,k)\Rightarrow
[a\cdot\alpha,a\cdot\beta]=-\alpha\cdot y\beta\,k\;,
\end{array}
$$
with $y^T=-y$ and $x^{lT}=x^l$. Finally
$$
\begin{array}{l}
J(h\cdot\alpha,a\cdot\beta,s_i)\Rightarrow
yR_i+R_i^Ty=0\;,\quad x^lR_i+R_i^Tx^l=f_{ij}{}^lx^j\;,
\\
J(h\cdot\alpha,h\cdot\beta,a\cdot\gamma)\Rightarrow
{1\over2}\alpha\,(\beta\cdot y\gamma)
  +R_l\alpha\,(\beta\cdot x^l\gamma)
  -(\alpha\leftrightarrow\beta)=\gamma\,(\alpha\cdot y\beta)\;.
\end{array}
$$
The trace of this `completeness relation' yields
$$
x^lR_l-(x^lR_l)^T+(2k+1)y=0\;.
$$

\subsection{The Invariant Scalar Product}

The simple Lie algebra $g$ has the invariant scalar product (unique up to an
overall factor)
$$
<X,Y>_g={1\over c_g}\Tr_g(\Ad(X)\,\Ad(Y))\;,\quad X,Y\in g\;,
$$
similarly the invariant scalar product on $sl(2)$ is
$$
<X,Y>_{sl(2)}={1\over c_{sl(2)}}\Tr_{sl(2)}(\Ad(X)\,\Ad(Y))\;,\quad
X,Y\in sl(2)\;,
$$
where the action of the adjoint representation is
$$
\Ad(X)\left|Y\right>=\left|[X,Y]\right>\;,\quad X,Y\in g\;.
$$
In order to compute the scalar product for $sl(2)$ we need the commutators
$$
\begin{array}{llll}
{}[d,[d,e]]=e\;,&
{}[d,[d,d]]=0\;,&
{}[d,[d,k]]=k&
\Rightarrow<d,d>_{sl(2)}={2\over c_{sl(2)}}\;,
\\
{}[e,[k,e]]=2e\;,&
{}[e,[k,d]]=2d\;,&
{}[e,[k,k]]=0&
\Rightarrow<e,k>_{sl(2)}={4\over c_{sl(2)}}\;.
\end{array}
$$
Choosing $c_{sl(2)}=2$ we obtain for $X=\epsilon e+\delta d+\kappa k$
$$
<X,X>_{sl(2)}=\delta^2+4\epsilon\kappa=
\delta^2+(\epsilon+\kappa)^2-(\epsilon-\kappa)^2\;.
$$
Extending the trace over all generators of $g$
$$
\begin{array}{lll}
{}[d,[d,h\cdot\alpha]]={1\over4}h\cdot\alpha\;,&
{}[d,[d,a\cdot\alpha]]={1\over4}a\cdot\alpha&
\Rightarrow<d,d>_g={k+2\over c_g}\;,
\\
{}[e,[k,h\cdot\alpha]]=h\cdot\alpha\;,&
{}[e,[k,a\cdot\alpha]]=0&
\Rightarrow<e,k>_g={2(k+2)\over c_g}\;.
\end{array}
$$
We choose $c_g=k+2$ such that the scalar product $<\cdot,\cdot>_{sl(2)}$
coincides with the restriction of $<\cdot,\cdot>_g$ to the subalgebra
$sl(2)$. The restriction of $<\cdot,\cdot>_g$ to the subalgebra $\bg$
similarly defines a particular invariant scalar product on $\bg$ (among
possibly different ones when $\bg$ is not simple).
For the scalar product $<s_i,s_j>$ we need the commutators
$$
[s_i,[s_j,h\cdot\alpha]]=h\cdot R_iR_j\alpha\;,\quad
[s_i,[s_j,a\cdot\alpha]]=a\cdot R_iR_j\alpha\;,
$$
and find
$$
<s_i,s_j>={\Tr_{\bg}(\Ad(s_i)\,\Ad(s_j))+2\Tr(R_i\,R_j)\over k+2}\;.
$$
For the scalar product $<h\cdot\alpha,a\cdot\beta>$ we need the commutators
$$
\begin{array}{l}
{}[h\cdot\alpha,[a\cdot\beta,e]]=(\alpha\cdot y\beta)\,e\;,\\
{}[h\cdot\alpha,[a\cdot\beta,h\cdot\gamma]]=
  h\cdot R_l\alpha\,(\beta\cdot x^l\gamma)
  -{1\over2}h\cdot\alpha\,(\beta\cdot y\gamma)\;,\\
{}[h\cdot\alpha,[a\cdot\beta,d]]=
  {1\over2}(\alpha\cdot y\beta)\,d+{1\over2}(\alpha\cdot x^l\beta)\,s_l\;,\\
{}[h\cdot\alpha,[a\cdot\beta,s_i]]=
  -(\alpha\cdot yR_i\beta)\,d-(\alpha\cdot x^jR_i\beta)\,s_j\;,\\
{}[h\cdot\alpha,[a\cdot\beta,a\cdot\gamma]]=
  -a\cdot\alpha\,(\beta\cdot y\gamma)\;,\\
{}[h\cdot\alpha,[a\cdot\beta,k]]=0\;.
\end{array}
$$
Collecting all terms we find
$$
<h\cdot\alpha,a\cdot\beta>={1\over k+2}( 3(\alpha\cdot y\beta)
  -\alpha\cdot x^lR_l\beta+\beta\cdot x^lR_l\alpha)\;,
$$
and finally using the completeness relation
$$
<h\cdot\alpha,a\cdot\beta>=2(\alpha\cdot y\beta)\;.
$$

\subsection{The Involutive Automorphism}

As noted above the two involutive automorphisms $\tau$ and $\tau'$ coincide
when restricted to the subalgebra $\bg\oplus sl(2)$, whereas
$\tau(X)=-\tau'(X)$ for $X=h\cdot\alpha+a\cdot\beta$. The action of $\tau$
on the subalgebra $sl(2)$ is given above, the action on $\bg$ can be
expressed by a matrix $T=(T^i{}_j)$
$$
\tau(s_i)=s_jT^j{}_i\;,\quad{\rm with}\quad
T^i{}_jT^j{}_k=\delta^i_k\;.
$$
From the commutators $[d,h_i]={1\over2}h_i$ and $[d,a_i]=-{1\over2}a_i$
together with $\tau(d)=-d$ we can deduce
$$
\tau(h\cdot\alpha)=a\cdot Z\alpha\;,\quad
\tau(a\cdot\alpha)=h\cdot Z^{-1}\alpha\;,
$$
with some nonsingular $2k\times2k$ matrix $Z$. The properties of this matrix
$Z$ can be determined by consistency requirements:
$$
\begin{array}{ll}
\begin{array}{l}
\tau([e,a\cdot\alpha])=[\tau(e),\tau(a\cdot\alpha)]\\
\tau([k,h\cdot\alpha])=[\tau(k),\tau(h\cdot\alpha)]
\end{array}&
\Rightarrow Z^{-1}=-Z\;,
\\
\begin{array}{l}
\tau([s_i,a\cdot\alpha])=[\tau(s_i),\tau(a\cdot\alpha)]\\
\tau([s_i,h\cdot\alpha])=[\tau(s_i),\tau(h\cdot\alpha)]
\end{array}&
\Rightarrow ZR_iZ=-\tau(R_i)\equiv-R_jT^j{}_i\;,
\\
\begin{array}{l}
\tau([h\cdot\alpha,a\cdot\beta])=[\tau(h\cdot\alpha),\tau(a\cdot\beta)]
\end{array}&
\Rightarrow Z^TyZ=y\;,\quad Z^Tx^kZ=T^k{}_lx^l\;.
\end{array}
$$

Note that the symmetric matrix $yZ$ is positive definite, since the scalar
product $<X,\tau'(X)>$ is negative definite. Changing the basis of the
representation space for $\brho$ yields modified matrices $R_i$, $Z$, $y$,
and $x^l$
$$
R_i\mapsto S^{-1}R_iS\;,\quad
Z\mapsto S^{-1}ZS\;,\quad
y\mapsto S^TyS\;,\quad
x^l\mapsto S^Tx^lS\;.
$$
We choose a basis such that $yZ=c\bfE_{2k}$, where $\bfE_n$ is the
$n$-dimensional unit matrix, and $c$ is the positive constant introduced in
Chapter~\ref{c_redu}, with the consequence
$$
\tau(R_i)=-R_i^T\;.
$$
The matrix $Z$ is then antisymmetric, i.e.
antihermitian and has therefore $k$ pairs of complex conjugate eigenvectors
with eigenvalues $\pm i$. This allows us to further specifiy the basis such
that
$$
Z=\left(\begin{array}{cc}0&-\eta^T\\\eta&0\end{array}\right)\;,\quad
y=c\left(\begin{array}{cc}0&\eta^T\\-\eta&0\end{array}\right)=-c\bY^{-1}\;,
\quad
\eta^T\eta=\bfE_k\;,
$$
and to identify the Lie algebra elements $k$, $a\cdot\alpha$, $d$, and $s_i$
as generators of infinitesimal twist gauge, electromagnetic gauge, scale,
and $\bG$ transformations respectively. We will occasionally decompose the
$2k$-dimensional vectors $A$, $a$, $h$, $\alpha$, \dots\ into $k$-dimensional
`electric' and `magnetic' parts $A_{(e)}$, $a_{(e)}$, \dots\ and $A_{(m)}$,
$a_{(m)}$, \dots\ respectively (with $A_{(e)}=B$ and $A_{(m)}=C$).

\boldsection{Parametrization of the $\sigma$-Model}
\label{c_sigma}

\subsection{The General Structure}

We use an `Iwasawa type' (KAN) decomposition of the Lie Algebra $g$ with the
two subgroups $H=\exp K$ and $T=\exp A\,\exp N$ and coset representatives
$$
\pi:G/H \mapsto T\;,
$$
(`triangular gauge'). The action of the group $G$ on these coset
representatives is
$$
G\ni u:\pi(x)\mapsto v(x)\pi(x)u^{-1}\;,\quad{\rm with}\quad
v(x)\in H\;.
$$
The transformation with the constant parameter $u\in G$ is combined
with a gauge transformation with parameter $v(x)\in H$ chosen such that
the transformed coset representative is again an element of the triangular
subgroup $T$. The infinitesimal form of these transformations is
$$
g\ni\delta g:\delta\pi(x)=\delta h(x)\pi(x)-\pi(x)\delta g\;,
\quad{\rm with}\quad\delta h(x)\in h\;.
$$
It is convenient to rewrite the variation of $\pi(x)$ in the form
$$
\delta\pi(x)=(\delta h(x)-\pi(x)\delta g\pi^{-1}(x))\pi(x)\;,
$$
that allows to determine $\delta h(x)\equiv\delta h(\pi(x),\delta g)$. Note
that the Iwasawa decomposition and the triangular subgroup $T$ are very
convenient but are not really needed; all we need are unique coset
representatives $\pi$ that allow us to compute $\delta h(\pi(x),\delta g)$.
For some of the coset spaces $\bG/\bH$ discussed in Appendix~\ref{a_models}
we will actually not use the triangular subgroup.

Given $\pi(x)$ we can compute two Lie algebra valued 1-forms $\cA$ and $\cJ$
$$
d\pi\,\pi^{-1}=\cA+\cJ=(\cA_m+\cJ_m)dx^m\;,\quad{\rm where}\quad
\tau(\cA)=\cA\;,\quad \tau(\cJ)=-\cJ\;,
$$
with transformation laws
$$
\delta\cJ=[\delta h,\cJ]\;,\quad
\delta\cA=d\delta h+[\delta h,\cJ]\;,
$$
i.e., $\cJ$ is a covariant (`matter') field and $\cA$ can be interpreted as
gauge connection. The $\sigma$-model Lagrangian is
$$
L=-{1\over4}<\cJ,\cJ>\;,
$$
with the implied contraction of 1-forms
$\cJ_mdx^m\,\cJ_ndx^n\mapsto\cJ_m\cJ^m$.
We can rewrite the resulting field equations
$$
\cD_m\cJ^m\equiv\nabla_m\cJ^m-[\cA_m,\cJ^m]=0\;,
$$
in the form
$$
\cD_m\cJ^m=\pi\nabla_mJ^m\pi^{-1}\;,\quad{\rm with}\quad
J=\pi^{-1}\cJ\pi\;,\quad L=-{1\over4}<J,J>\;,
$$
and express $J$ in terms of $\pi$ as
$$
J={1\over2}\mu^{-1}d\mu\;,\quad{\rm with}\quad
\mu=\tau(\pi^{-1})\pi\;,
$$
and with the linear transformation laws
$$
G\ni u:\mu\mapsto \tau(u)\mu u^{-1}\;,\quad
J\mapsto uJu^{-1}\;.
$$

\boldsubsection{The $SL(2)/SO(2)$ $\sigma$-Model}

Dimensional reduction of gravity from 4 to 3 dimensions yields the well
known $SL(2)/SO(2)$ $\sigma$-model with two scalar fields, the square
$\Delta$ of the Killing vector and the twist potential $\psi$.

The coset representatives $\tdpi\in\tdG\equiv SL(2)$ are
$$
\tdpi=e^{\ln\Delta\,d}\,e^{\psi\,k}\;.
$$
We will need the expressions
$$
\delta\tdpi\,\tdpi^{-1}=
  {\delta\Delta\over\Delta}\,d+{\delta\psi\over\Delta}\,k\;,\quad
\tdpi^{-1}\delta\tdpi=
  {\delta\Delta\over\Delta}\,d+(\delta\psi-\psi{\delta\Delta\over\Delta})\,k\;,
$$
in order to compute the variation of the fields $\Delta$ and $\psi$ under
infinitesimal transformations generated by $e$ (Ehlers transformation), $d$
(scale transformation), and $k$ (twist gauge transformation). The
expression $\tdpi\delta\tdg$ has already the required form for
$\delta\tdg=k$ or $d$ (the generators of the triangular subgroup $T$)
and therefore the `compensating' gauge transformation vanishes in these two
cases:
$$
\begin{array}{llll}
\delta\tdg=k:&
\tdpi^{-1}\delta\tdpi=-k&
{}\Rightarrow \delta\Delta=0\;,& \delta\psi=-1\;,
\\
\delta\tdg=d:&
\tdpi^{-1}\delta\tdpi=-d&
{}\Rightarrow \delta\Delta=-\Delta\;,& \delta\psi=-\psi\;.
\end{array}
$$
For the Ehlers transformation we find
$$
\delta\tdg=e:\quad
\delta\tdpi\,\tdpi^{-1}=\delta\tdh-\tdpi\,e\,\tdpi^{-1}
=\delta\tdh-\Delta\,e+2\psi\,d+{\psi^2\over\Delta}k\;.
$$
Since the coefficient of $e$ in $\delta\tdpi\,\tdpi^{-1}$ has to
vanish we must choose $\delta\tdh=\Delta(e-k)$ and obtain
$$
\delta\tdg=e:\quad
\delta\tdpi\,\tdpi^{-1}=2\psi\,d+({\psi^2\over\Delta}-\Delta)k\quad
{}\Rightarrow \delta\Delta=2\psi\Delta\;,\quad\delta\psi=\psi^2-\Delta^2\;.
$$
The commutators of these variations of the fields $\Delta$ and $\psi$
reflect the structure of the Lie algebra
$$
[\delta(X),\delta(Y)]=\delta([Y,X])\;,
$$
where $\delta(X)$ are the variations under the transformation generated by
$X\in\tdg$.

The decomposition of $d\tdpi\,\tdpi^{-1}$ yields
$$
\tdcA={d\psi\over2\Delta}(k-e)\;,\quad
\tdcJ={d\Delta\over\Delta}\,d+{d\psi\over2\Delta}(e+k)\;,\quad
L=-{(\partial\Delta)^2+(\partial\psi)^2\over4\Delta^2}\;,
$$
and the gauge invariant current is
$$
\tdJ=\tdpi^{-1}\tdcJ\tdpi=
{d\psi\over2\Delta^2}\,e+{\Delta d\Delta+\psi d\psi\over\Delta^2}\,d
+{(\Delta^2-\psi^2)d\psi-2\psi\Delta d\Delta\over2\Delta^2}\,k\;.
$$

We can easily translate these relations for Lie algebra and group elements
into matrix equations, using the 2-dimensional matrix representation
$\tdrho$
$$
\tdrho(\epsilon e+\delta d+\kappa k)=\left(\begin{array}{cc}
   {1\over2}\delta&\epsilon\\\kappa&-{1\over2}\delta\end{array}\right)\;,
\quad
<X,Y>=2\Tr(\tdrho(X)\,\tdrho(Y))\;,
$$
with
$$
\tdrho(\tdtau(X))=-\tdrho^T(X)=\tdy\tdrho(X)\tdy^{-1}\;,
\quad
\tdy=i\sigma_2=\left(\begin{array}{cc}0&1\\-1&0\end{array}\right)\;,
$$
and $\tdrho(\tdpi)=\tdP$, $\tdrho(\tdmu)=\tdM$,
$$
\tdP=\left(\begin{array}{cc}
   \sqrt\Delta&0\\
   {\psi\over\sqrt\Delta}&{1\over\sqrt\Delta}\end{array}\right)\;,
\quad
\tdM=\tdP^T\tdP=\left(\begin{array}{cc}
   \Delta+\Delta^{-1}\psi^2&\Delta^{-1}\psi\\
   \Delta^{-1}\psi&\Delta^{-1}\end{array}\right)\;.
$$

\boldsubsection{The $G/H$ $\sigma$-Model}
\label{s_sigma}

The fields of the $G/H$ (or $G/H'$) $\sigma$-model are the scalars from the
4-dimensional $\bG/\bH$ $\sigma$-model described by coset
representatives $\bpi$, the two scalars $\Delta$ and $\psi$ from
4-dimensional gravity and $2k$ scalar (`electromagnetic') potentials $A^i$
from the $k$ vector fields. Note that he twist vector $\omega_m$ now has a
contribution from the electromagnetic potentials
$$
\omega=\omega_mdx^m=d\psi-{1\over2}A\cdot ydA\;.
$$

We choose $G/H$ coset representatives $\pi$ in the triangular group
$T\subset G$
$$
\pi=e^{\ln\Delta\,d}\,\bpi\,e^{a\cdot A+\psi\,k}\;,
$$
with the $\bG/\bH$ coset representatives $\bpi\in\bT\subset\bG$. We will
again need the expressions
$$
\begin{array}{l}
\delta\pi\,\pi^{-1}=
{\delta\Delta\over\Delta}\,d+\delta\bpi\,\bpi^{-1}
+{a\cdot\bP\delta A\over\sqrt\Delta}
+{\delta\psi-{1\over2}A\cdot y\delta A\over\Delta}\,k\;,
\\
\pi^{-1}\delta\pi=
{\delta\Delta\over\Delta}\,d+\bpi^{-1}\delta\bpi
+a\cdot(\delta A+\bP^{-1}\delta\bP A-A{\delta\Delta\over2\Delta})
\\\qquad\qquad
+(\delta\psi+{1\over2}A\cdot y\delta A
+{1\over2}A\cdot y\bP^{-1}\delta\bP A
-\psi{\delta\Delta\over\Delta})\,k\;,
\end{array}
$$
where $\bP$ represents the group element $\bpi\in\bG$ in the
$2k$-dimensional matrix representation $\brho$
$$
\bg\ni s_i\mapsto R(s_i)=R_i\;,\quad
\bG\ni\bpi\mapsto\bP\;,\quad{\rm with}\quad
\bpi\,h\cdot\alpha\,\bpi^{-1}=h\cdot\bP\alpha\;.
$$
For the infinitesimal transformations generated by $k$ (shift of the twist
potential), $a$ (shift of the electromagnetic potentials), and $d$ (scale
transformation) we have $\delta h=0$ and thus $\pi^{-1}\delta\pi=-\delta g$.
The variations of the fields are:
$$
\begin{array}{lllll}
\delta g=k:&
\delta\Delta=0\;,&
\delta\bpi=0\;,&
\delta A=0\;,&
\delta\psi=-1\;,
\\
\delta g=a\cdot\alpha:&
\delta\Delta=0\;,&
\delta\bpi=0\;,&
\delta A=-\alpha\;,&
\delta\psi={1\over2}A\cdot y\alpha\;,
\\
\delta g=d:&
\delta\Delta=-\Delta\;,&
\delta\bpi=0\;,&
\delta A=-{1\over2}A\;,&
\delta\psi=-\psi\;.
\end{array}
$$
For $\delta g=\delta\bg\in\bg$ we need the compensating gauge
transformation $\delta h=\delta\bh(\bpi,\delta\bg)$ resulting in
$$
\begin{array}{ll}
\delta g=\delta\bg:&
\pi^{-1}\delta\pi=\pi^{-1}\delta\bh\pi-\delta\bg
\\&\quad
=\bpi^{-1}\delta\bh\bpi-\delta\bg
+a\cdot R(\bpi^{-1}\delta\bh\bpi)A
+{1\over2}(A\cdot yR(\bpi^{-1}\delta\bh\bpi)A)\,k\;.
\end{array}
$$
Comparing coefficients and using
$\bP^{-1}\delta\bP=R(\bpi^{-1}\delta\bpi)$ we find
$$
\delta\Delta=0\;,\quad
\delta\bpi=\delta\bh\bpi-\bpi\delta\bg\;,\quad
\delta A=R(\delta\bg)A\;,\quad
\delta\psi=0\;.
$$

In order to determine the compensating gauge transformation $\delta h$
required for the infinitesimal transformations generated by $h$ (Harrison
transformations) and $e$ (Ehlers transformation) we first compute
$\pi\,\delta g\,\pi^{-1}$. For the Harrison transformations with
$\delta g=h\cdot\alpha$ we obtain
$$
\begin{array}{l}
\delta h-\pi\,\delta g\,\pi^{-1}=
\delta h-\exp({a\cdot\bP A\over\sqrt\Delta}+{\psi\over\Delta}k)
\,h\cdot\bP\alpha\,\sqrt\Delta\,
\exp-({a\cdot\bP A\over\sqrt\Delta}+{\psi\over\Delta}k)
\\\qquad
=\delta h-h\cdot\bP\alpha\,\sqrt\Delta
-A\cdot y\alpha\,d
+A\cdot x^l\alpha\,\bpi s_l\bpi^{-1}
\\\qquad\qquad
+{1\over\sqrt\Delta}\,a\cdot\bP(\alpha\,\psi
-{1\over4} A\,A\cdot y\alpha
-{1\over2}R_lA\,A\cdot x^l\alpha)
\\\qquad\qquad
-{\psi\over\Delta}\,A\cdot y\alpha\,k
+{1\over6\Delta}\,A\cdot yR_lA\,A\cdot x^l\alpha\,k\;.
\end{array}
$$
This requires
$$
\delta h(\pi,h\cdot\alpha)=
(h\cdot\bP\alpha+a\cdot Z\bP\alpha)\,\sqrt\Delta
-A\cdot x^l\alpha\,\delta\bh(\bpi,s_l)\;,
$$
and yields
$$
\begin{array}{l}
\delta\pi\,\pi^{-1}=
-A\cdot y\alpha\,d
-A\cdot x^l\alpha\,(\delta\bh(\bpi,s_l)-\bpi\,s_l\,\bpi^{-1})
+a\cdot Z\bP\alpha\,\sqrt\Delta
\\\qquad\qquad
+{1\over\sqrt\Delta}\,a\cdot\bP(\alpha\,\psi
-{1\over4} A\,A\cdot y\alpha
-{1\over2}R_lA\,A\cdot x^l\alpha)
\\\qquad\qquad
-{\psi\over\Delta}\,A\cdot y\alpha\,k
+{1\over6\Delta}\,A\cdot yR_lA\,A\cdot x^l\alpha\,k\;.
\end{array}
$$
Comparing coefficients we find the variations of the fields
$$
\begin{array}{l}
\delta\Delta=-A\cdot y\alpha\,\Delta\;,
\\
\delta\bpi=-A\cdot x^l\alpha\,
  (\delta\bh(\bpi,s_l)\,\bpi-\bpi\,s_l)\;,
\\
\delta A=
\alpha\,\psi
+cy^{-1}\bM\alpha\,\Delta
-{1\over4}A\,A\cdot y\alpha
-{1\over2}R_lA\,A\cdot x^l\alpha\;,
\\
\delta\psi=
-{1\over2}A\cdot y\alpha\,\psi
+{c\over2}A\cdot\bM\alpha\,\Delta
-{1\over12}A\cdot yR_lA\,A\cdot x^l\alpha\;,
\end{array}
$$
with the positive definite symmetric matrix $\bM=\bP^T\bP$.

For the Ehlers transformations with $\delta g=e$ we obtain
$$
\begin{array}{l}
\delta h-\pi\,\delta g\,\pi^{-1}=
\delta h-\exp({a\cdot\bP A\over\sqrt\Delta}+{\psi\over\Delta}k)
\,\Delta\,e\,
\exp-({a\cdot\bP A\over\sqrt\Delta}+{\psi\over\Delta}k)
\\\qquad
=\delta h-e\,\Delta
-h\cdot\bP A\,\sqrt\Delta
+2\psi\,d
+{1\over2}A\cdot x^lA\,\bpi s_l\bpi^{-1}
\\\qquad\qquad
+{1\over\sqrt\Delta}\,a\cdot\bP(A\,\psi
-{1\over6}R_lA\,A\cdot x^lA)
\\\qquad\qquad
+{\psi^2\over\Delta}\,k
+{1\over24\Delta}\,A\cdot yR_lA\,A\cdot x^lA\,k\;.
\end{array}
$$
This requires
$$
\delta h(\pi,e)=
\Delta(e-k)
+(h\cdot\bP A+a\cdot Z\bP A)\,\sqrt\Delta
-{1\over2}A\cdot x^lA\,\delta\bh(\bpi,s_l)\;,
$$
and yields
$$
\begin{array}{l}
\delta\pi\,\pi^{-1}=
2\psi\,d
-{1\over2}A\cdot x^lA\,(\delta\bh(\bpi,s_l)-\bpi\,s_l\,\bpi^{-1})
+a\cdot Z\bP A\,\sqrt\Delta
\\\qquad\qquad
+{1\over\sqrt\Delta}\,a\cdot\bP(A\,\psi
-{1\over6}R_lA\,A\cdot x^lA)
\\\qquad\qquad
+{\psi^2-\Delta^2\over\Delta}\,k
+{1\over24\Delta}\,A\cdot yR_lA\,A\cdot x^lA\,k\;.
\end{array}
$$
Comparing coefficients we find the variations of the fields
$$
\begin{array}{l}
\delta\Delta=2\psi\,\Delta\;,
\\
\delta\bpi=-{1\over2}A\cdot x^lA\,
  (\delta\bh(\bpi,s_l)\,\bpi-\bpi\,s_l)\;,
\\
\delta A=
A\,\psi
+cy^{-1}\bM A\,\Delta
-{1\over6}R_lA\,A\cdot x^lA\;,
\\
\delta\psi=
\psi^2-\Delta^2
+{c\over2}A\cdot\bM A\,\Delta
-{1\over24}A\cdot yR_lA\,A\cdot x^lA\;.
\end{array}
$$

The decomposition of $d\pi\,\pi^{-1}$ yields
$$
\begin{array}{l}
\cA=\bcA
+{1\over2\sqrt\Delta}(a\cdot\bP\,dA-h\cdot Z\bP\,dA)
+{\omega\over2\Delta}(k-e)\;,
\\
\cJ={d\Delta\over\Delta}\,d+\bcJ
+{1\over2\sqrt\Delta}(a\cdot\bP\,dA+h\cdot Z\bP\,dA)
+{\omega\over2\Delta}(e+k)\;,
\end{array}
$$
where $\bcA$ and $\bcJ$ are the quantities derived from $\bpi$.
The Lagrangian is
$$
L=-{1\over4}<\cJ,\cJ>
=-{1\over4}<\bcJ,\bcJ>
+{c\over4\Delta}\partial A\cdot\bM\,\partial A
-{(\partial\Delta)^2+\omega^2\over4\Delta^2}\;.
$$
The gauge invariant current is
$$
J=\pi^{-1}\cJ\pi=N^{-1}\hJ\,N\;,\quad
N=\exp(a\cdot A+\psi\,k)\;,
$$
with
$$
\hJ={\omega\over2\Delta^2}e
+{1\over2\Delta}h\cdot Z\bM\,dA
+{d\Delta\over\Delta}\,d+\bJ
+{1\over2}a\cdot dA
+{\omega\over2}k\;.
$$
In order to compute $N^{-1}\hJ\,N$ let us first evaluate $N^{-1}X\,N$ for
all $X\in g$
$$
\begin{array}{l}
N^{-1}e\,N=e
-h\cdot A
+2\psi\,d
-{1\over2}A\cdot x^iA\,s_i
\\\qquad\qquad
-a\cdot A\,\psi
-{1\over6}a\cdot R_iA\,A\cdot x^iA
-\psi^2\,k
-{1\over24}A\cdot x^iA\,A\cdot yR_iA\,k\;,
\\
N^{-1}h\cdot\alpha\,N=h\cdot\alpha
-A\cdot y\alpha\,d
+A\cdot x^i\alpha\,s_i
+a\cdot\alpha\,\psi
+{1\over4}a\cdot A\,A\cdot y\alpha
\\\qquad\qquad
+{1\over2}a\cdot R_iA\,A\cdot x^i\alpha
+A\cdot y\alpha\,\psi\,k
+{1\over6}A\cdot x^iA\,A\cdot yR_i\alpha\,k\;,
\\
N^{-1}d\,N=d-{1\over2}a\cdot A-\psi\,k\;,
\\
N^{-1}s_i\,N=s_i+a\cdot R_iA+{1\over2}A\cdot yR_iA\,k\;,
\\
N^{-1}a\cdot\alpha\,N=a\cdot\alpha+A\cdot y\alpha\,k\;,
\\
N^{-1}k\,N=k\;,
\end{array}
$$
Expanding $J$ with respect to the generators
$$
J={1\over2}J^ee+{1\over2}h\cdot ZJ^h
+J^dd+J^is_i
+{1\over2}a\cdot J^a+{1\over2}J^kk\;,
$$
and collecting all terms we obtain the conserved currents
$$
\newcommand{\D}[1]{\displaystyle\mathop{#1}\limits^{}}
\begin{array}{l}
\D{J^e=
{\omega\over\Delta^2}}\;,
\\
\D{J^h=
{1\over\Delta}\bM\,dA
+ZA{\omega\over\Delta^2}}\;,
\\
\D{J^d=
{d\Delta\over\Delta}
-{c\over2\Delta}A\cdot\bM\,dA
+{\omega\over\Delta^2}\psi}\;,
\\
\D{J^is_i=
\bJ
+{1\over2\Delta}A\cdot x^iZ\bM\,dA\,s_i
-{\omega\over4\Delta^2}A\cdot x^iA\,s_i}\;,
\\
\D{J^a=
dA
-A{d\Delta\over\Delta}
+2R(\bJ)A
+{1\over2\Delta}R_iA\,A\cdot x^iZ\bM\,dA
+{1\over\Delta}Z\bM\,dA\,\psi}
\\\qquad
\D{+{c\over4\Delta}A\,A\cdot\bM\,dA
-{\omega\over\Delta^2}A\,\psi
-{\omega\over6\Delta^2}R_iA\,A\cdot x^iA}\;,
\\
\D{J^k=
\omega
+A\cdot ydA
-2{d\Delta\over\Delta}\,\psi
+A\cdot yR(\bJ)A
+{c\over\Delta}A\cdot\bM\,dA\,\psi}
\\\qquad
\D{-{c\over6\Delta}A\cdot x^iA\,A\cdot R_i^T\bM\,dA
-{\omega\over\Delta^2}\psi^2
-{\omega\over24\Delta^2}A\cdot x^iA\,A\cdot yR_iA}\;,
\end{array}
$$
and the Lagrangian
$$
L=-{1\over4}<J,J>
 =-{1\over4}\left(J^iJ^j<s_i,s_j>-cJ^a\cdot J^h
    +J^dJ^d+J^eJ^k\right)\;.
$$
Using the identity $x^i<s_i,s_j>=-2yR_j$ it is straight forward, although
tedious, to verify that this reproduces the expression given above.

\section{Application to Black Holes}
\label{c_black}

In our previous paper with Gibbons~\cite{BMG} we have used the
$\sigma$-model formulation in order to generalize a number of results on
stationary black holes of the Einstein-Maxwell theory to a large class of
4-dimensional theories with abelian gauge fields and scalars. With the more
detailed knowledge of their group theoretical structure gained in this
paper, some further applications are conceivable. Here we just give simple
derivations of two such results for strictly stationary single black holes.
First we generalize a quadratic mass formula of Heusler~\cite{Heusler} to
all the theories under consideration. It is quite remarkable, how the use of
the group theoretical structure simplifies the derivation. As a second
application we state the action of the Ehlers-Harrison transformation on the
charges of the black hole.

Consider a strictly stationary single black hole solution for one of the
theories discussed in this article in its dimensionally reduced form.
We can choose suitable coordinates on $\Sigma_3$ as well as a suitable gauge
such that the behaviour of the fields at infinity is
$$
h_{mn}(x)=\delta_{mn}+O({1\over r})\;,\quad
\pi(x)=\exp({\pi_1\over r}+O({1\over r^2}))\;.
$$
The element $\pi_1\in t$ (where $t$ is the Lie algebra of the triangular
group $T$) contains the charges
$$
\pi_1=-2md+2nk+a\cdot q+\bpi_1\;,
$$
the total mass $m$, the NUT-charge $n$, the combined electric and magnetic
charges $q$, and scalar charges determined by $\bpi_1\in\bt$.

Integrating the conserved currents $J$ over a closed surface $\Sigma$
enclosing the black hole yields the Lie algebra valued charge
$$
Q={1\over4\pi}\oint J^md\Sigma_m\;.
$$
Evaluating this integral at infinity we obtain
$$
Q={1\over2}(\tau(\pi_1)-\pi_1)
   =2md-n(e+k)-{1\over2}(a\cdot q+h\cdot Zq)+Q^is_i\;,
$$
with the scalar charges $\bQ=Q^is_i={1\over2}(\tau(\bpi_1)-\bpi_1)$.

Due to the assumption of strict stationarity, the field $\Delta$ vanishes on
the horizon $\cH$ and is positive everywhere outside $\cH$. The boundary
conditions at the horizon~\cite{Carter} imply that $\psi$, $A$, and the
surface gravity $\kappa={1\over2}\sqrt{h^{mn}\partial_m\Delta\partial_n\Delta}$
are constant on the horizon, whereas the scalars fields $\phi^i$ are finite but
may vary. Moreover, the twist vector $\omega_m$ vanishes linearly with
$\Delta$, and the area of the horizon is given by
$$
a_\cH=\lim_{\Delta\to0}{1\over\Delta}
   \oint\limits_{\Delta=\Const}n^md\Sigma_m\;,
$$
where $n$ is the unit normal to the surface $\Delta=\Const$ with respect to
the rescaled metric $h_{mn}$.

\subsection{A Quadratic Mass Formula}

Evaluating the integral for $Q$ on the horizion yields
$$
Q=N_\cH^{-1}Q_\cH N_\cH\;,
$$
where $N_\cH$ is the value of $N=\exp(a\cdot A+\psi k)$ on the horizon, and
the charges $Q_\cH$ on the horizon are
$$
Q_\cH={1\over4\pi}\oint_\cH\hJ^md\Sigma_m
   =2m_\cH d-n_\cH e-{1\over2}h\cdot Zq_\cH\;,
$$
with the `irreducible mass'
$$
m_\cH={1\over8\pi}\oint_\cH{\partial_m\Delta\over\Delta}d\Sigma^m
   ={1\over4\pi}\kappa_\cH a_\cH\;,
$$
and the NUT and vector charges `at the horizon'
$$
n_\cH=-{1\over8\pi}\oint_\cH{\omega_m\over\Delta^2}d\Sigma^m\;,\quad
q_\cH=-{1\over4\pi}\oint_\cH{\bM\partial_mA\over\Delta}d\Sigma^m\;.
$$
Note that $Q_\cH$ has no contributions with the generators $s_i$, $a$, and
$k$, because the corresponding terms in $\hJ$ carry no inverse powers of
$\Delta$.

Using the invariance of the scalar product, the relation between $Q$ and
$Q_\cH$ implies $<Q,Q>=<Q_\cH,Q_\cH>$. Furthermore the structure of the
scalar product (pairing $e$ with $k$, etc.)\ yields $<Q_\cH,Q_\cH>=4m_\cH^2$
and hence the desired quadratic mass formula
$$
m^2+n^2-{c\over4}q^Tq+<\bQ,\bQ>_{\bg}=m_\cH^2\;,
$$
extending the well known relation $m^2+n^2-q^2-p^2=m_\cH^2$ for the
generalized Reissner-Nordstr{\o}m solutions with electric and magnetic
charges $q$ and $p$ to a large class of theories.

\boldsubsection{The Action of $G$ on Black Holes}

The action of $G$, given explicitly in Section~\ref{s_sigma} for the
infinitesimal transformations, preserves the strict stationarity as well as
the boundary conditions on the horizon, and leaves the irreducible mass
invariant. The action of $G$ on the conserved currents is
$$
G\ni u:J\mapsto uJu^{-1}\;.
$$

In order to preserve the boundary conditions at infinity, we have to
restrict the transformations to the subgroup $H$, with the action on the Lie
algebra valued charge $Q$ given by
$$
H\ni u:Q\mapsto uQu^{-1}\;.
$$
For the infinital transformation generated by $\delta g=e-k$ this yields
$$
\delta m=-2n\;,\quad
\delta n=2m\;,\quad
\delta q=Zq\;,\quad
\delta\bQ=0\;,
$$
for the transformation generated by $\delta g=h\cdot\alpha+a\cdot Z\alpha$
$$
\begin{array}{ll}
\displaystyle\delta m=-{c\over2}q\cdot Z\alpha\;,&
\displaystyle\delta q=2\alpha n-2Z\alpha m+2R(\bQ)Z\alpha\;,
\\
\displaystyle\delta n={c\over2}q\cdot\alpha\;,&
\displaystyle\delta\bQ=-{1\over2}q\cdot x^l\alpha(s_l-\tau(s_l))\;,
\end{array}
$$
and for the transformations generated by $\delta g=\delta\bg\in\bh$
$$
\delta m=0\;,\quad
\delta n=0\;,\quad
\delta q=R(\delta\bg)q\;,\quad
\delta\bQ=[\delta\bg,\bQ]\;.
$$

Starting from a nondegenerate black hole (i.e.\ $m_\cH>0$) with charges $m$,
$n$, $q$, and $\bQ$, we can apply a suitable group element from $H$ to
obtain a new black hole with the same $m_\cH$, and with charges $n=q=0$.
Using the identity $Q=N_\cH^{-1}Q_\cH N_\cH$ we conclude $n_\cH=q_\cH=0$,
and therefore $m=m_\cH$, $\bQ=\psi_\cH=A_\cH=0$.

\appendix

\section[The Individual Models]{Appendix: The Individual Models}
\label{a_models}

Dimensional reduction of 4-dimensional theories consisting of a $\bG/\bH$
$\sigma$-model coupled to $k$ abelian vector fields and gravity to
3~dimensions yields $(\Dim\bG-\Dim\bH)+2k+2$ scalars coupled to gravity.
The conditions that all these scalars form one $G/H$ $\sigma$-model have
been discussed in~\cite{BMG}. In Table~\ref{t_list}, reproduced from that
paper, we list 15 different possibilities, all with a simple Lie group $G$.
In the following we discuss some of these cases in detail and finally
indicate a general procedure applicable to all cases.

\boldsubsection{$SL(n+2)/SO(n,2)$}
\label{s_sl}

These are the Ka{\l}uza-Klein theories obtained from pure gravity in $D=n+4$
dimensions by dimensional reduction with respect to $n$ commuting space like
Killing vectors. The 4-dimensional theory consists of $k=n$ vector fields
and a $GL(n)/SO(n)$ $\sigma$-model with ${1\over2}n(n+1)$ scalars. Let
$\hpi$ be the $GL(n)/SO(n)$ coset representative and $\hrho$ the
$k$-dimensional matrix representation
$$
\hrho:\hpi\mapsto\hrho(\hpi)=\hP\;,\quad
\tau(\hP)=\hP^{T-1}\;,\quad
\hrho(s_i)=r_i\;.
$$
The action for the 4-dimensional theory is given by
$$
<\bJ,\bJ>_{\bg}=2\Tr(\hJ\,\hJ)-{2\over n+2}\Tr(\hJ)\Tr(\hJ)\;,\quad
c=1\;,\quad
\tdmu=\hM\;,\quad
\tdnu=0\;,
$$
where the symmetric matrix $\hM=\hP^T\hP$ contains the (rescaled) scalar
products of the $n$ Killing vectors, and the $\sigma$-model currents are
$$
\hrho(\bJ)=\hJ={1\over2}\hM^{-1}d\hM\;.
$$
The theory is invariant under the action of $GL(n)$ and $SO(n)$
$$
GL(n)\ni u:
B_a\mapsto u\,B_a\;,\quad
\hP\mapsto v\,\hP\,u^{-1}\;,\quad{\rm with}\quad v\in SO(n)\;.
$$

Further dimensional reduction with respect to a time like Killing vector
yields (with a suitable choice for $\eta$) the 3-dimensional theory with
$$
\bM=\left(\begin{array}{cc}\hM&0\\0&\hM^{-1}\end{array}\right)\;,
\quad
R_i=\left(\begin{array}{cc}r_i&0\\0&-r_i^T\end{array}\right)\;,\quad
y=\left(\begin{array}{cc}0&\bfE_n\\-\bfE_n&0\end{array}\right)\;.
$$
This model has the unique property that the scalar potentials $A$ transform
under a reducible representation of the group $\bG=GL(n)$, due to the fact
that the action of $\bG$ on the field strengths $B_{ab}$ involves no duality
transformations; this leads to an unambiguous decomposition into electric
and magnetic potentials $A_{(e)}$ and $A_{(m)}$ respectively.

In order to exhibit the $SL(n+2)/SO(n,2)$ $\sigma$-model consider the
$(n+2)$-dimensional matrix representation $\rho(X)$ of the element
$X=\epsilon e+h\cdot\alpha+\delta d+\lambda^is_i+a\cdot\beta+\kappa k$
of the Lie algebra $sl(n+2)$ and the involutive automorphism
$\rho(\tau(X))=-D^{-1}\rho^T(X)D$ with
$$
\rho(X)=\left(\begin{array}{ccc}
  {1\over2}\delta&-\alpha_{(m)}^T&\epsilon\\
  \beta_{(e)}&\lambda^ir_i&\alpha_{(e)}\\
  \kappa&\beta_{(m)}^T&-{1\over2}\delta
  \end{array}\right)
-{\Tr(\lambda^ir_i)\over n+2}\bfE_{n+2}\;,
\quad
D=\left(\begin{array}{ccc}1&0&0\\0&-\bfE_n&0\\0&0&1\end{array}\right)\;,
$$
and with the scalar product $<X,Y>=2\Tr(\rho(X)\,\rho(Y))$. Finally we
obtain
$$
P=\rho(\pi)=(\det\hP)^{-{1\over n+2}}
\left(\begin{array}{ccc}
   \sqrt\Delta&0&0\\
   \hP A_{(e)}&\hP&0\\
   {1\over\sqrt\Delta}(\psi+{1\over2}A_{(m)}^TA_{(e)})&
   {1\over\sqrt\Delta}A_{(m)}^T&{1\over\sqrt\Delta}
   \end{array}\right)\;,
$$
and $\rho(J)={1\over2}M^{-1}dM$ with $M=P^TDP$.

\boldsubsection{$SU(p+1,q+1)/S(U(p,1)\times U(1,q))$}
\label{s_su}

These are 4-dimensional theories with $k=p+q$ vector fields and the $2pq$
real scalars from $U(p,q)/(U(p)\times U(q))$. For $q=0$
there are no scalars, the only effect of the $U(p)/U(p)$ $\sigma$-model is
the action of $U(p)$ on the field strength. For $p=1$, $q=0$ this is the
Einstein-Maxwell theory; the theories with $p>1$, $q=0$ are generalizations
with several vector fields.

Let $\bpi$ be the $U(p,q)/(U(p)\times U(q))$ coset representative, $\hrho$
the $k$-di\-men\-sional complex matrix representation
$$
\hrho:\bpi\mapsto\hrho(\bpi)=\hP\;,\quad
\hV\hP=(\hP^+)^{-1}\hV\;,\quad
\tau(\hP)=(\hP^+)^{-1}\;,
$$
with the real $U(p,q)$ metric $\hV=\hV^T=\hV^*=\hV^{-1}$, and $\brho$ the
corresponding $2k$-dimensional real representation
$$
\brho:\bpi\mapsto\brho(\bpi)=\bP=
\left(\begin{array}{rr}\Re\hP&-\Im\hP\\\Im\hP&\Re\hP\end{array}\right)\;.
$$
Decomposing the hermitian matrix $\hM=\hP^+\hP$ satisfying $\hM\hV\hM=\hV$
into real and imaginary parts $\tdM=\tdM^T=\Re\hM$ and $\tdN=-\tdN^T=\Im\hM$
we obtain the relations
$$
\tdM\hV\tdM-\tdN\hV\tdN=\hV\;,\quad
\tdM\hV\tdN+\tdN\hV\tdM=0\;.
$$
This allows us to express the action for the 4-dimensional theory with $c=4$
in terms of the symmetric matrices
$$
\tdmu=\hV\tdM^{-1}\hV\;,\quad
\tdnu=\hV\tdM^{-1}\tdN\;,
$$
and the invariant scalar product on $u(p,q)$
$$
<X,Y>_{\bg}=2\Tr(\hrho(X)\,\hrho(Y))-{2\over p+q+2}\Tr\hrho(X)\Tr\hrho(Y)\;.
$$
Note, however, that $\det\hM=1$ and therfore $\Tr\hrho(J)=0$.

Choosing $\eta=\hV$, we obtain the the 3-dimensional theory with
$$
\bM=\left(\begin{array}{rr}\tdM&-\tdN\\\tdN&\tdM\end{array}\right)\;,\quad
y=c\left(\begin{array}{cc}0&\hV\\-\hV&0\end{array}\right)\;.
$$
In order to construct the $(k+2)$-dimensional matrix representation $\rho$
of $SU(p+1,q+1)$ we first rearrange the $2k$ real components of $A$ into a
$k$-dimensional complex vector $\bfA=A_{(e)}+iA_{(m)}$ and similarly for
$a$, $h$, $\alpha$, etc. The element
$X=\epsilon e+h\cdot\alpha+\delta d+\lambda^is_i+a\cdot\beta+\kappa k$
of $su(p+1,q+1)$ is represented by
$$
\rho(X)=\left(\begin{array}{ccc}
   {1\over2}\delta&-\sqrt2\bfalpha^+\hV&\epsilon\\
   i\sqrt2\bfbeta&\lambda^i\hrho(s_i)&i\sqrt2\bfalpha\\
   \kappa&\sqrt2\bfbeta^+\hV&-{1\over2}\delta
   \end{array}\right)-{\Tr(\lambda^i\hrho(s_i))\over k+2}\bfE_{k+2}\;,
$$
such that $\rho(X)=-V^{-1}\rho^+(X)V$ and
$\rho(\tau(X))=-D^{-1}\rho^+(X)D$ with
$$
V=\left(\begin{array}{ccc}
   0&0&-i\\0&\hV&0\\i&0&0
   \end{array}\right)\;,\quad
D=\left(\begin{array}{ccc}
   1&0&0\\0&-\bfE_k&0\\0&0&1
   \end{array}\right)\;,
$$
and with the scalar product $<X,Y>=2\Tr(\rho(X)\,\rho(Y))$. Finally we
obtain
$$
P=\rho(\pi)=(\det\hP)^{-{1\over k+2}}
\left(\begin{array}{ccc}
   \sqrt\Delta&0&0\\
   i\sqrt2\hP\bfA&\hP&0\\
   {1\over\sqrt\Delta}(\psi+i\bfA^+\hV\bfA)&
   {\sqrt2\over\sqrt\Delta}\bfA^+\hV&{1\over\sqrt\Delta}
   \end{array}\right)\;,
$$
the twist vector $\omega=d\psi-2\Im\bfA^+\hV d\bfA$, and
$\rho(J)={1\over2}M^{-1}dM$ with $M=P^+DP$.

\boldsubsection{$SO(p+2,q+2)/(SO(p,2)\times SO(2,q))$}
\label{s_so}

These are 4-dimensional theories with $k=p+q$ vector fields, with $pq$
scalars from $SO(p,q)/(SO(p)\times SO(q))$, and with a
dilaton $\varphi$ and an axion $\chi$ from $SL(2)/SO(2)$.
Some well known examples are the Einstein-Maxwell-dilaton-axion theory with
$p=1$, $q=0$~\cite{Galtsov}, the bosonic sector of $N=4$ supergravity with
$p=6$, $q=0$, and the (dimensionally reduced) bosonic sector of
10-dimensional supergravity with $p=q=6$.

Let $\hpi$ be the $SO(p,q)/(SO(p)\times SO(q))$ coset representative and
$\hrho$ the $k$-dimensional matrix representation
$$
\hrho:\hpi\mapsto\hrho(\hpi)=\hP\;,\quad
\hV\hP=\hP^{T-1}\hV\;,\quad
\tau(\hP)=\hP^{T-1}\;,
$$
with the real $SO(p,q)$ metric $\hV=\hV^T=\hV^{-1}$, and consider the action for
the 4-dimensional theory with
$$
<\bJ,\bJ>_{\bg}={(\partial\varphi)^2+(\partial\chi)^2\over\varphi^2}
   +\Tr(\hJ\,\hJ)\;,\quad
c=1\;,\quad
\tdmu=\varphi\hM\;,\quad
\tdnu=\chi\hV\;,
$$
where
$$
\hM=\hP^T\hP\;,\quad
\hJ={1\over2}\hM^{-1}d\hM\;.
$$
Choosing $\eta=\hV$, we obtain the 3-dimensional theory with
$\bM=\hM\otimes\tdM$, $y=\hV\otimes\tdy$, $\tdy=i\sigma_2$, where
$$
\tdP=\left(\begin{array}{cc}
   \sqrt\varphi&0\\
   {\chi\over\sqrt\varphi}&{1\over\sqrt\varphi}\end{array}\right)\;,
\quad
\tdM=\tdP^T\tdP=\left(\begin{array}{cc}
   \varphi+\varphi^{-1}\chi^2&\varphi^{-1}\chi\\
   \varphi^{-1}\chi&\varphi^{-1}\end{array}\right)\;,
$$
parametrizes the $SL(2)/SO(2)$ $\sigma$-model.

In order to construct the $(k+4)$-dimensional matrix representation $\rho$
of $SO(p+2,q+2)$ we first rearrange the $2k$ components of $A$ into a
$k\times2$ matrix $\bfA=(A_{(e)},A_{(m)})$ and similarly for $a$, $h$,
$\alpha$, \dots\ and split the generators $s_i$ of $\bG$ into
$\hs_i\in\hg=so(p,q)$ and $\tds_i\in\tdg=sl(2)$. The element
$X=\epsilon e+h\cdot\alpha+\delta d+\lambda^is_i+a\cdot\beta+\kappa k$
of $so(p+2,q+2)$ is represented by
$$
\rho(X)=\left(\begin{array}{ccc}
   {1\over2}\delta\bfE_2+\tdlambda^i\tdrho(\tds_i)&
      -\bfalpha^T\hV&\epsilon\bfE_k\\
   \bfbeta\tdy&\hlambda^i\hrho(\hs_i)&\bfalpha\tdy\\
   \kappa\bfE_2&\bfbeta^T\hV&
      -{1\over2}\delta\bfE_2+\tdlambda^i\tdrho(\tds_i)
   \end{array}\right)\;,
$$
such that $\rho(X)=-V^{-1}\rho^T(X)V$ and
$\rho(\tau(X))=-D^{-1}\rho^T(X)D$ with
$$
V=\left(\begin{array}{ccc}
   0&0&\tdy\\0&\hV&0\\-\tdy&0&0
   \end{array}\right)\;,\quad
D=\left(\begin{array}{ccc}
   \bfE_2&0&0\\0&-\bfE_k&0\\0&0&\bfE_2
   \end{array}\right)\;,
$$
and with the scalar product $<X,Y>=\Tr(\rho(X)\,\rho(Y))$. Finally we obtain
$$
P=\rho(\pi)=\left(\begin{array}{ccc}
   \Delta^{1/2}\tdP&0&0\\
   \hP\bfA\tdy&\hP&0\\
   \Delta^{-1/2}(\psi\bfE_2+{1\over2}\bfA^T\hV\bfA\tdy)&
   \Delta^{-1/2}\bfA^T\hV&\Delta^{-1/2}\tdP
   \end{array}\right)\;,
$$
and $\rho(J)={1\over2}M^{-1}dM$ with $M=P^TDP$.

\boldsubsection{$SO^*(2n+4)/U(n,2)$}
\label{s_so*}

These are 4-dimensional theories with $k=2n$ vector fields and the $n(n-1)$
scalars from $SO^*(2n)/U(n)\times SU(2)/SU(2)$. In the
following we assume $n>1$, since for $n=1$ this is the $SU(3,1)/U(2,1)$
theory already discussed in Section~\ref{s_su}. The $4n$ electromagnetic
potentials transform under the $2n$-dimensional representation $\hrho$ of
$SO^*(2n)$ and under the 2-dimensional representation $\tdrho$ of $SU(2)$,
and we will therefore split the generators $s_i$ of $\bG$ into
$\hs_i\in\hg=so^*(2n)$ and $\tds_i\in\tdg=su(2)$.

The group $SO^*(2n)$ is defined as the subgroup of $SO(2n;\bfC)$ that leaves
the antihermitian form $\varphi^+\,i\sigma_2\otimes\bfE_n\,\chi$
invariant. Allowing for a change of basis we obtain
$$
\hrho:\hs_i\mapsto\hR_i\;,\quad
\hV\hR_i+\hR^+_i\hV=0\;,\quad
\hW\hR_i+\hR^T_i\hW=0\;,\quad
\tau(\hR_i)=-\hR^+_i\;,
$$
with matrices $\hV=-\hV^+$ and $\hW=\hW^T$ such that
$\hC\equiv\hV^{-1}\hW^+=\hW^{-1}\hV^T$. This implies the reality condition
$\hR_i=\hC\hR^*_i\hC^{-1}$ with $\hC\hC^*=-\bfE_{2n}$. Hence the
representation $\hrho$ is `pseudo real', i.e., is equivalent to its complex
conjugate but cannot be written with real matrices; the same holds true for
the  2-dimensional representation $\tdrho$ of $su(2)$ with
$$
\tdrho:\tds_i\mapsto\tdR_i={i\over2}\sigma_i=\tau(\tdR_i)
   =-\tdR^+_i=-\tdy^{-1}\tdR^T_i\tdy\;,\quad
\tdy=i\sigma_2=\left(\begin{array}{cc}0&1\\-1&0\end{array}\right)\;,
$$
and $\tdR_i=\tdy\tdR^*_i\tdy^{-1}$.

We can now construct the $(2n+4)$-dimensional matrix representation $\rho$
of $so^*(2n+4)$ with the Lie algebra element
$X=\epsilon e+h\cdot\alpha+\delta d+\lambda^is_i+a\cdot\beta+\kappa k$
represented by
$$
\rho(X)=\left(\begin{array}{ccc}
   {1\over2}\delta\bfE_2+\tdlambda^i\tdrho(\tds_i)&
      -\bfalpha^+\hV&\epsilon\bfE_2\\
   \bfbeta&\hlambda^i\hrho(\hs_i)&\bfalpha\\
   \kappa\bfE_2&\bfbeta^+\hV&
      -{1\over2}\delta\bfE_2+\tdlambda^i\tdrho(\tds_i)
   \end{array}\right)\;,
$$
such that $V\rho(X)+\rho^+(X)V=0$, $W\rho(X)+\rho^T(X)W=0$, and
$\rho(\tau(X))=-D^{-1}\rho^+(X)D$ with
$$
V=\left(\!\begin{array}{ccc}
   0&0&-\bfE_2\\0&\hV&0\\\bfE_2&0&0
   \end{array}\!\right)\,,\;
W=\left(\!\begin{array}{ccc}
   0&0&\tdy\\0&\hW&0\\-\tdy&0&0
   \end{array}\!\right)\,,\;
D=\left(\!\begin{array}{ccc}
   \bfE_2&0&0\\0&-\bfE_{2n}&0\\0&0&\bfE_2
   \end{array}\!\right)\,,
$$
and with the scalar product $<X,Y>=\Tr(\rho(X)\,\rho(Y))$. The
representation matrices satisfy the reality condition
$\rho(X)^*=C^{-1}\rho(X)C$ with the `charge conjugation' matrix
$$
C=V^{-1}W^+=W^{-1}V^T=\left(\begin{array}{ccc}
   -\tdy&0&0\\0&\hC&0\\0&0&-\tdy
   \end{array}\right)\;,
$$
The $2n\times2$ matrix $\bfA$ (and similarly $\bfalpha$, $\bfbeta$, \dots)
satisfies the reality condition $\bfA=\hC\bfA^*\tdy$ and can be expressed
in terms of the $4n$ components of $A$, the precise form depending on the
choice of $\hV$ and $\hW$. Choosing $\hV=i\sigma_1\otimes\bfE_n$ and
$\hW=\sigma_3\otimes\bfE_n$ with $\hC=-\sigma_2\otimes\bfE_n$, as well as
$\bfA=(A_{(c)},\hC A_{(c)}^*)$ with $A_{(c)}=A_{(e)}+iA_{(m)}$, we obtain
$$
(\alpha\cdot y\beta)\bfE_2=
   \bfbeta^+\hV\bfalpha-\bfalpha^+\hV\bfbeta\;,\quad
y=\left(\begin{array}{cc}
   0&2\sigma_1\otimes\bfE_n\\-2\sigma_1\otimes\bfE_n&0
   \end{array}\right)\;,
$$
i.e.\ $c=2$ and $\eta=\sigma_1\otimes\bfE_n$, as well as the
$4n$-dimensional real representation $\brho=\hrho\otimes\tdrho$ of $\bg$ in
the form
$$
\brho:\lambda^is_i\mapsto
   \hlambda^i\left(\!\begin{array}{rr}
      \Re\hR_i&-\Im\hR_i\\\Im\hR_i&\Re\hR_i
      \end{array}\!\right)
   +\left(\!\begin{array}{cc}
      {\tdlambda^1\over2}i\sigma_2&
      {\tdlambda^2\over2}i\sigma_2+{\tdlambda^3\over2}\bfE_2\\
      {\tdlambda^2\over2}i\sigma_2-{\tdlambda^3\over2}\bfE_2&
      -{\tdlambda^1\over2}i\sigma_2
      \end{array}\!\right)\otimes\bfE_n\;.
$$
Finally we obtain
$$
P=\rho(\pi)=\left(\begin{array}{ccc}
   \Delta^{1/2}\bfE_2&0&0\\
   \hP\bfA&\hP&0\\
   \Delta^{-1/2}(\psi\bfE_2+{1\over2}\bfA^+\hV\bfA)&
   \Delta^{-1/2}\bfA^+\hV&\Delta^{-1/2}\bfE_2
   \end{array}\right)\;,
$$
and $\rho(J)={1\over2}M^{-1}dM$ with $M=P^+DP$, and therefore
$$
<J,J>=\Tr(\rho(J)\rho(J))=
   \Tr(\hrho(\hJ)\hrho(\hJ))
   -{2\over\Delta}\partial A^+_{(c)}\hM\partial A_{(c)}
   +{(\partial\Delta)^2+\omega^2\over\Delta^2}\;,
$$
with $\hrho(\hJ)={1\over2}\hM^{-1}d\hM$, $\hM=\hP^+\hP$, and
$\hM\,\sigma_1\otimes\bfE_n\,\hM=\sigma_1\otimes\bfE_n$. We can therefore
proceed as in Section~\ref{s_su}, and express the 4-dimensional theory in
terms of the symmetric matrices
$$
\tdmu=\sigma_1\otimes\bfE_n\,(\Re\hM)^{-1}\,\sigma_1\otimes\bfE_n\;,\quad
\tdnu=\sigma_1\otimes\bfE_n\,(\Re\hM)^{-1}\,\Im\hM\;.
$$

\boldsubsection{$Sp(2n+2;\bfR)/U(n,1)$}
\label{s_sp}

These correspond to 4-dimensional theories with $k=n$ vector fields and the
$n(n+1)$ scalars from $Sp(2n;\bfR)/U(n)$. For $n=1$ this
is again the Einstein-Maxwell-dilaton-axion theory~\cite{Galtsov}.

Consider the action for the 4-dimensional theory with
$$
<\bJ,\bJ>_{\bg}=
   \Tr(\hM^{-1}\partial\hM\hM^{-1}\partial\hM)
   +\Tr(\hM^{-1}\partial\hN\hM^{-1}\partial\hN)\;,
$$
$c=1$, $\tdmu=\hM$, $\tdnu=\hN$, where the symmetric matrices $\hM=\hP^T\hP$
and $\hN$ describe scalar `dilaton' fields and pseudo scalar `axion' fields
respectively. Choosing $\eta=\bfS_n$, where $\bfS_n$ is the $n$-dimensional
`skew diagonal unit matrix' with elements $(\bfS_n)_{ij}=\delta_{i+j,n+1}$,
we obtain the 3-dimensional theory with $\bM=\bP^T\bP$, and
$$
\bP=\left(\begin{array}{cc}\hP&0\\\bfS_n\hP^{T-1}\hN&\bfS_n\hP^{T-1}\bfS_n
   \end{array}\right)\;,\quad
y=\left(\begin{array}{cc}0&\bfS_n\\-\bfS_n&0
   \end{array}\right)\;.
$$
The matrix $\bP$ is the $Sp(2n;\bfR)/U(n)$ coset representative $\bpi$ in
the $2n$-dimensional matrix representation
$$
\brho:\bpi\mapsto\brho(\bpi)=\bP\;,\quad
y\bP=\bP^{T-1}y\;,\quad
\tau(\bP)=\bP^{T-1}\;,
$$
with the `symplectic metric' $y=-y^T=-y^{-1}$, and we can express the
invariant scalar product on the Lie algebra $sl(2n;\bfR)$ in the form
$<X,Y>_{\bg}=2\Tr(\brho(X)\,\brho(Y))$.

In order to exhibit the $Sp(2n+2;\bfR)/U(n,1)$ $\sigma$-model consider the
$(2n+2)$-dimensional matrix representation $\rho(X)$ of the element
$X=\epsilon e+h\cdot\alpha+\delta d+\lambda^is_i+a\cdot\beta+\kappa k$
of the Lie algebra $sp(2n+2;\bfR)$
$$
\rho(X)=\left(\begin{array}{ccc}
  {1\over2}\delta&{1\over\sqrt2}\alpha^Ty&\epsilon\\
  {1\over\sqrt2}\beta&\lambda^i\brho(s_i)&{1\over\sqrt2}\alpha\\
  \kappa&-{1\over\sqrt2}\beta^Ty&-{1\over2}\delta
  \end{array}\right)\;,
$$
with the symplectic metric $V$ such that
$\rho(X)=-V^{-1}\rho^T(X)V$ and with the involutive automorphism
$\rho(\tau(X))=-D^{-1}\rho^T(X)D$, where
$$
V=\left(\begin{array}{cc}0&\bfS_{n+1}\\-\bfS_{n+1}&0
   \end{array}\right)\;,
\quad
D=\left(\begin{array}{ccc}1&0&0\\0&-\bfE_{2n}&0\\0&0&1\end{array}\right)\;,
$$
and with the scalar product $<X,Y>=2\Tr(\rho(X)\,\rho(Y))$. Finally we
obtain
$$
P=\rho(\pi)=\left(\begin{array}{ccc}
   \sqrt\Delta&0&0\\
   {1\over\sqrt2}\bP A&\bP&0\\
   {1\over\sqrt\Delta}\psi&
   -{1\over\sqrt{2\Delta}}A^Ty&{1\over\sqrt\Delta}
   \end{array}\right)\;,
$$
and $\rho(J)={1\over2}M^{-1}dM$ with $M=P^TDP$.

\subsection{The General Procedure}
\label{s_general}

For the remaining cases of Table~\ref{t_list} the group $G$ is one of the
exceptional groups $G_2$, $F_4$, $E_6$, $E_7$, or $E_8$. Some of these cases
describe the bosonic sector of supergravity theories, e.g., $N=8$
supergravity in 4~dimensions with $\bG/\bH=E_{7(+7)}/SU(8)$ and
$G/H=E_{8(+8)}/SO^*(16)$~\cite{Markus}.
Instead of discussing each of these cases in
detail, we outline the general procedure and apply it to Case~6 with
$G/H=G_{2(+2)}/(SU(1,1)\times SU(1,1))$, $\bG/\bH=SU(1,1)/U(1)$, and $k=2$
vector fields.

Given the $G/H$ and $\bG/\bH$ $\sigma$-models in 3~and 4~dimensions, we know
that the 4-dimensional theory has
$k={1\over2}(\dim(G/H)-\dim(\bG/\bH)-2)={1\over4}(\dim G-\dim\bG-3)$ vector
fields. Parametrizing the Lie algebra of $G$ as described in
Chapter~\ref{c_lie}, and choosing some $c>0$ we find that the
electromagnetic potentials transform with a $2k$-dimensional real
representation $\brho$ of $\bG$ with
$$
\brho:s_i\mapsto\brho(s_i)=R_i=-y^{-1}R_i^Ty\;,\quad
y=c\left(\begin{array}{cc}0&\eta^T\\-\eta&0\end{array}\right)\;,
$$
where $\eta$ is an arbitrary orthogonal matrix and $\tau(R_i)=-R_i^T$.
The representation matrices have therefore the structure
$$
R_i=\left(\begin{array}{cc}
   A_i&B_i\eta^T\\\eta C_i&-\eta A_i^T\eta^T
   \end{array}\right)\;,
$$
with symmetric $k\times k$ matrices $B$ and $C$. In our example with
$\bG/\bH=SU(1,1)/U(1)\sim SL(2)/SO(2)$ we choose $c=1$ and generators $\bd$,
$\be$, and $\bk$ of $\bG$ with the commutation relations
$$
[\bd,\be]=\be\;,\quad [\bd,\bk]=-\bk\;,\quad [\be,\bk]=2\bd\;,
$$
the automorphism
$$
\tau(\be)=-\bk\;,\quad \tau(\bd)=-\bd\;,\quad \tau(\bk)=-\be\;,
$$
and the 4-dimensional matrix representation of the Lie algebra element
$X=\bepsilon\be+\bdelta\bd+\bkappa\bk$
$$
\brho(X)=\left(\begin{array}{cccc}
   {3\over2}\bdelta&\sqrt3\bepsilon&0&0\\
   \sqrt3\bkappa&{1\over2}\bdelta&2\bepsilon&0\\
   0&2\bkappa&-{1\over2}\bdelta&\sqrt3\bepsilon\\
   0&0&\sqrt3\bkappa&-{3\over2}\bdelta\\
   \end{array}\right)\;,\quad
\eta=\left(\begin{array}{cc}0&1\\-1&0\end{array}\right)\;.
$$

Next we compute the invariant scalar product $<s_i,s_j>$, in our
example $<X,X>=3\bdelta^2+12\bepsilon\bkappa$, and use the identity
$x^i<s_i,s_j>=-2yR_j$ to determine the matrices $x^i$.

We would like to choose coset representatives $\bpi$ from a triangular group
$\bT$, and choose a basis for the representation $\brho$, such that the
generators of $\bT$ are represented by matrices $R_i$ with $B_i=0$ (and
$A_i$ suitably restricted). If that is possible, we obtain matrices
$\bP=\brho(\bpi)$ and $\bM=\brho(\bmu)$
$$
\bP=\left(\begin{array}{cc}
   \tdP&0\\\eta\tdP^{T-1}\tdnu&\eta\tdP^{T-1}\eta^T
   \end{array}\right)\;,\quad
\bM=\bP^T\bP=\left(\begin{array}{cc}
   \tdmu+\tdnu\tdmu^{-1}\tdnu&\tdnu\tdmu^{-1}\eta^T\\
   \eta\tdmu^{-1}\tdnu&\eta\tdmu^{-1}\eta^T
   \end{array}\right)\;,
$$
with symmetric matrices $\tdmu=\tdP^T\tdP$ and $\tdnu$, and can then
reconstruct the 4-dimensional theory in terms of these matrices. All we
really need is, however, a way to express the matrices $\bM$ in terms of
$\tdmu$, $\tdnu$, and $\eta$. In our example we choose
$\bpi=e^{\ln\varphi\bd}e^{\chi \bk}$ represented by
$$
\bP=\left(\begin{array}{cccc}
   \varphi^{3/2}&0&0&0\\
   \sqrt3\varphi^{1/2}\chi&\varphi^{1/2}&0&0\\
   \sqrt3\varphi^{-1/2}\chi^2&2\varphi^{-1/2}\chi&\varphi^{-1/2}&0\\
   \varphi^{-3/2}\chi^3&\sqrt3\varphi^{-3/2}\chi^2&
      \sqrt3\varphi^{-3/2}\chi&\varphi^{-3/2}
   \end{array}\right)\;,
$$
and finally obtain
$$
\tdmu=\left(\begin{array}{cc}
   \varphi^3+3\varphi\chi^2&\sqrt3\varphi\chi\\\sqrt3\varphi\chi&\varphi
\end{array}\right)\;,\quad
\tdnu=\left(\begin{array}{cc}
  2\chi^3&\sqrt3\chi^2\\\sqrt3\chi^2&2\chi
\end{array}\right)\;,
$$
and
$$
<\bJ,\bJ>=3{(\partial\varphi)^2+(\partial\chi)^2\over\varphi^2}\;.
$$

In the following we will show for the remaining Cases~7--15 from
Table~\ref{t_list}, that the matrices $\bM$ can indeed be expressed in terms
of $\tdmu$, $\tdnu$, and $\eta$ as required. In most cases this can be
achieved by choosing $\bP$ in the form given above. Assume that there exist
bases in the Lie algebra $\bg$ and in the representation space for $\brho$
such that for each $R_i$ at least one of the submatrices $B_i$ and $C_i$ is
zero. We can then choose the triangular subgroup $\bT$ for the coset
representatives, generated by Lie algebra elements $s_i$ with $B_i=0$ and
with $A_i$ suitably restricted.  This is always possible if there exists a
`coset generator' $X=-\tau(X)\in\bg$ such that all eigenvalues of $\brho(X)$
are nonzero. The symmetric matrix $\brho(X)$ can be diagonalized by an
orthogonal transformation, and we can choose the eigenvectors with positive
and negative eigenvalues as first and last~$k$ elements respectively of a
new basis in the representation space for $\brho$, automatically preserving
the structure of $y$. Next we can diagonalize $\Ad X$, which is symmetric
provided we choose a basis for $\bg$ with $<s_i,\tau(s_j)>=-\delta_{ij}$,
such that $[X,s_i]=\xi_is_i$. In these new bases $B_i=0$ except when
$\xi_i>0$ and $C_i=0$ except when $\xi_i<0$ as desired. In the following we
demonstrate the existence of such an $X$ for each of the Cases~7--9
and~11--15 of Table~\ref{t_list}. For Case~10, where no such $X$ can be
found, we will directly demonstrate how to express the matrices $\bM$ in
terms of symmetric matrices $\tdmu$ and $\tdnu$.

\boldsubsubsection{The 14-Dimensional Representation of $Sp(6;\bfR)$}
\label{ss_repc3}

In Case~7 of Table~\ref{t_list} there are 7~vector fields and the
electromagnetic potentials transform under one of the two inequivalent
14-dimensional representations of $Sp(6;\bfR)$. Decomposing this
representation with respect to the subgroup $SL(2)\otimes SL(2)\otimes
SL(2)$ (using the isomorphism $Sp(2;\bfR)=SL(2)$) yields
$2\otimes2\otimes2\oplus 2\otimes1\otimes1\oplus 1\otimes2\otimes1\oplus
1\otimes1\otimes2$. Each of the three $SL(2)$ subgroups has a coset
generator $\bd_i$ with eigenvalues $\pm{1\over2}$ in the 2-dimensional real
representation denoted by~2. Therefore, we can choose $X=\bd_1+\bd_2+\bd_3$
with eigenvalues $\pm{1\over2}$ and $\pm{3\over2}$.

\boldsubsubsection{The 20-Dimensional Representation of $A_5$}
\label{ss_repa5}

In Cases~8--10 of Table~\ref{t_list} there are 10~vector fields and the
electromagnetic potentials transform under the 20-dimensional representation
of one of the noncompact forms $sl(6)$, $su(3,3)$, or $su(5,1)$ of the Lie
algebra $A_5$. As representation space we may take the totally antisymmetric
3-index tensors $\varphi_{ijk}$, real for $sl(6)$ and subject to the reality
condition
$\bvarphi^{ijk}\equiv\gamma^{il}\gamma^{jm}\gamma^{kn}(\varphi_{lmn})^*=
{1\over6}\epsilon^{ijklmn}\varphi_{lmn}$ for $su(3,3)$ and $su(5,1)$, where
$\gamma^{ij}$ is the $su(p,q)$ metric (chosen as
$\gamma=\left(\begin{array}{cc}\bfE_p&0\\0&-\bfE_q\end{array}\right)$ for
simplicity).

For $sl(6)$ and $su(3,3)$ we can decompose the representation with respect to
the subalgebra $sl(2)\oplus sl(2)\oplus sl(2)$ as in
Subsection~\ref{ss_repc3} (using this time the isomorphism $su(1,1)=sl(2)$).
Distributing the three indices of $\varphi_{ijk}$ among the three $sl(2)$
subalgebras yields again the representations $2\otimes2\otimes2$,
$2\otimes1\otimes1$, $1\otimes2\otimes1$, and $2\otimes1\otimes1$.
Therefore, we can again choose $X=\bd_1+\bd_2+\bd_3$ with eigenvalues
$\pm{1\over2}$ and $\pm{3\over2}$.

This procedure does, however, not work for Case~10 with $su(5,1)$, and it
may even be impossible to bring the matrices $\bP$ into `block triagonal'
form. We therefore have to analyze the matrices $\bM$ in some detail. Let
$\hrho$ be the 6-dimensional complex representation of $su(5,1)$ with
$$
\hrho:s_i\mapsto\hrho(s_i)=\hR_i\;,\quad
\hV\hR_i+\hR^+_i\hV=0\;,\quad
\tau(\hR_i)=-\hR^+_i\;,
$$
with the $su(5,1)$ metric $\hV=\hV^+$; we choose
$$
\hV=\left(\begin{array}{cc}\bfE_5&0\\0&-1\end{array}\right)\;,\quad
\lambda^i\hR_i=\left(\begin{array}{cc}a&b\\b^+&-\Tr a\end{array}\right)\;,
$$
where $a=(a_i{}^j)$, $i,j=1,\ldots,5$ is (the matrix representative of) an
element of $u(5)$, and $b=(b_i)$, $b^+=(\bb^j)$ represent the coset
generators. Next consider the action of $su(5,1)$ on the totally
antisymmetric 3-index tensors $\varphi_{ijk}$, choosing
$\varphi_{ij}\equiv\varphi_{ij6}$ and $\bvarphi^{ij}=(\varphi_{ij})^*$ as
basis for this 20-dimensional representation
$$
\tdrho:s_i\mapsto\tdR_i\;,\quad
\tdV\tdR_i+\tdR^+_i\tdV=0\;,\quad
\tdW\tdR_i+\tdR^T_i\tdW=0\;,\quad
\tau(\tdR_i)=-\tdR^+_i\;,
$$
with matrices
$$
\lambda^i\tdR_i=\left(\!\begin{array}{cc}
   a_{ij}{}^{kl}&\bb_{ijkl}\\b^{ijkl}&\ba^{ij}{}_{kl}
   \end{array}\!\right),\quad
\tdV=\left(\!\begin{array}{cc}-i\bfE_{10}&0\\0&i\bfE_{10}
   \end{array}\!\right),\quad
\tdW=\left(\!\begin{array}{cc}0&i\bfE_{10}\\-i\bfE_{10}&0
   \end{array}\!\right),
$$
and matrix elements
$$
a_{ij}{}^{kl}=2a_{[i}{}^{[k}\delta_{j]}^{l]}
-\delta_{[i}^{[k}\delta_{j]}^{l]}a_m{}^m=-\ba^{kl}{}_{ij}\;,\quad
b^{ijkl}={1\over2}\epsilon^{ijklm}b_m=(\bb_{ijkl})^*\;.
$$
We finally obtain the real 20-dimensional representation $\brho$ of
$su(5,1)$ with $\Re\varphi_{ij}$ and $\Im\varphi_{ij}$ as new basis
$$
\brho:s_i\mapsto\brho(s_i)=R_i=S^{-1}\tdR_iS\;,\quad
\bV R_i+\ R^T_i\bV=0\;,\quad
\tau(R_i)=-R^T_i\;,
$$
with
$$
S={1\over\sqrt2}\left(\begin{array}{cc}1&i\\1&-i
   \end{array}\right)\;,\quad
\bV=S^+\tdV S=S^T\tdW S
   =\left(\begin{array}{cc}0&\bfE_{10}\\-\bfE_{10}&0\end{array}\right)\;,
$$
and therefore $\brho(\bpi)=\bP$ with $\bV\bP=\bP^{T-1}\bV$, and
$\brho(\bmu)=\bP^T\bP=\bM$ with $\bM\bV\bM=\bV$. The last equation implies
that $\bM$ has the required form with symmetric matrices $\tdmu$, $\tdnu$
and with $\eta=\bfE_{10}$.

\boldsubsubsection{The 32-Dimensional Representation of $D_6$}
\label{ss_repd6}

In Cases~11--13 of Table~\ref{t_list} there are 16~vector fields and the
electromagnetic potentials transform under a 32-dimensional real `spinor'
representation of one of the noncompact forms $so(6,6)$, $so^*(12)$, or
$so(10,2)$ of the Lie algebra $D_6$. The Lie algebra $D_n$ has two
inequivalent $2^{n-1}$-dimensional `chiral' spinor representations
$S^n_\pm$. For $so(p,q)$ with $p=q\pmod8$ they are both real, for $so(p,q)$
with $p=q+4\pmod8$ they are both pseudo real, for $so(p,q)$ with
$p=q\pm2\pmod8$ and for $so^*(4p+2)$ they are complex conjugate, whereas
$S^n_+$ is real and $S^n_-$ is pseudo real for $so^*(4p)$. Decomposing the
representations $S^{p+q}_\pm$ with respect to the subalgebra $D_p\oplus D_q$
yields $S^{p+q}_\pm=S^p_+\otimes S^q_\pm\oplus S^p_-\otimes S^q_\mp$.

For $so(6,6)$ and $so(10,2)$ we use the subalgebra $so(2,2)\oplus D_4$
with $D_4=so(4,4)$ and $D_4=so(8)$ respectively and obtain
$S^6_+=S^2_+\otimes S^4_+\oplus S^2_-\otimes S^4_-$ with real spinors
$S^2_\pm$ and $S^4_\pm$. Using the isomorphism $so(2,2)=sl(2)\oplus sl(2)$,
we find $S^2_+=2\otimes1$ and $S^2_-=1\otimes2$. Therefore, we can choose
$X=\bd_++\bd_-$ with eigenvalues $\pm{1\over2}$, where $\bd_\pm$ are the
coset generators of the two $sl(2)$ subalgebras.

For $so^*(12)$ we use the subalgebra $so^*(4)\oplus so^*(4)\oplus so^*(4)$
and obtain $S^6_+=\sum S^2_\pm\otimes S^2_\pm\otimes S^2_\pm$, where the sum
contains all combinations with an even number of $S^2_-$. Using the
isomorphism $so^*(4)=sl(2)\oplus su(2)$ we find $S^2_+=2\otimes1$ and
$S^2_-=1\otimes2'$, where $2'$ is the 2-dimensional pseudo real
representation of $su(2)$. Therefore, we can choose $X=\bd_1+\bd_2+\bd_2$
with eigenvalues $\pm{1\over2}$ and $\pm{3\over2}$, where $\bd_i$ are the
coset generators of the three $sl(2)$ subalgebras.

\boldsubsubsection{The 56-Dimensional Representation of $E_7$}
\label{ss_repe7}

In Cases~14 and~15 of Table~\ref{t_list} there are 28~vector fields and the
electromagnetic potentials transform under the 56-dimensional representation
of one of the noncompact forms $E_{7(+7)}$ or $E_{7(-25)}$ of the Lie
algebra $E_7$. We first decompose the representation with respect to the
subalgebra $sl(2)+D_6$, with $D_6=so(6,6)$ for $E_{7(+7)}$ or $D_6=so(10,2)$
for $E_{7(-25)}$, and obtain $2\otimes V^6\oplus1\otimes S^6_+$, where $V^6$
denotes the vector representation of $D_6$. Further decomposing the
representations of $D_6$ with respect to the subalgebra
$sl(2)\oplus sl(2)\oplus D_4$ as in Subsection~\ref{ss_repd6} finally yields
$V^6=2\otimes2\otimes1\oplus1\otimes1\otimes V^4$. Therefore, we can choose
$X=\bd_1+\bd_2+\bd_2$ with eigenvalues $\pm{1\over2}$ and $\pm{3\over2}$,
where $\bd_i$ are again the coset generators of the three $sl(2)$
subalgebras.

\filbreak 

\end{document}